\documentclass[submission, Phys]{SciPost}

\hypersetup{
    colorlinks,
    linkcolor={red!50!black},
    citecolor={blue!50!black},
    urlcolor={blue!80!black}
}

\usepackage[bitstream-charter]{mathdesign}
\DeclareSymbolFont{usualmathcal}{OMS}{cmsy}{m}{n}
\DeclareSymbolFontAlphabet{\mathcal}{usualmathcal}

\usepackage{units}
\usepackage{mathtools}
\usepackage{bm}
\usepackage{graphicx}
\usepackage{physics}
\usepackage{slashed}
\usepackage{ytableau}

\DeclareMathOperator{\sgn}{sgn}
\newcommand{\sld}[1]{\slashed{#1}}

\renewcommand{\b}[1]{\boldsymbol{#1}}
\renewcommand{\c}[1]{\mathcal{#1}}

\providecommand{\tabularnewline}{\\}

\begin{document}

\begin{center}{\large \textbf{
Monopoles in Dirac spin liquids and their symmetries from instanton calculus\\
}}\end{center}

\begin{center}
G. Shankar\textsuperscript{1} and
Joseph Maciejko\textsuperscript{1,2}
\end{center}

\begin{center}
{\bf 1} Department of Physics, University of Alberta, Edmonton, Alberta T6G 2E1, Canada
\\
{\bf 2} Theoretical Physics Institute, University of Alberta, Edmonton, Alberta T6G 2E1, Canada
\\
\end{center}

\begin{center}
\today
\end{center}


\section*{Abstract}
{\bf
The Dirac spin liquid (DSL) is a two-dimensional (2D) fractionalized Mott insulator featuring massless Dirac spinon excitations coupled to a compact $U(1)$ gauge field, which allows for flux-tunneling instanton events described by magnetic monopoles in (2+1)D Euclidean spacetime. The state-operator correspondence of conformal field theory has been used recently to define associated monopole operators and determine their quantum numbers, which encode the microscopic symmetries of conventional ordered phases proximate to the DSL. In this work, we utilize semiclassical instanton methods not relying on conformal invariance to construct monopole operators directly in (2+1)D spacetime as instanton-induced 't~Hooft vertices, i.e., fermion-number-violating effective interactions originating from zero modes of the Euclidean Dirac operator in an instanton background. In the presence of a flavor-adjoint fermion mass, resummation of the instanton gas is shown to select the correct monopole to be proliferated, in accordance with predictions of the state-operator correspondence. We also show that our instanton-based approach is able to determine monopole quantum numbers on bipartite lattices.
}

\vspace{10pt}
\noindent\rule{\textwidth}{1pt}
\tableofcontents\thispagestyle{fancy}
\noindent\rule{\textwidth}{1pt}
\vspace{10pt}

\section{Introduction}
\label{sec:intro}
A quantum spin liquid is a quintessential example of a fractionalized
phase in strongly correlated systems, whose low-energy description
is best afforded by a deconfined gauge theory~\cite{savary2016}.
The parton construction is a systematic approach to derive such a
description~\cite{wen2002,wen1991,wen2007}. In such an approach,
the lattice spins are rewritten as a composite of fermions or bosons
(partons) glued together by an emergent gauge field. While these partons
remain confined in conventional phases, a quantum spin liquid is characterized
by their deconfinement at low energy.

Of the various spin liquids that have been proposed, the Dirac spin liquid (DSL)
is of special renown for its candidate role as a ``parent state''
for several competing orders in two spatial dimensions (2D) on various
lattice geometries~\cite{hermele2005,hastings2000,ran2007,hermele2008,song2019a,song2020,xu2019}.
As known and reviewed below, a low-energy description of the DSL state
is afforded by compact quantum electrodynamics in three spacetime
dimensions (CQED$_{3}$) with $N_{f}\!=\!4$ flavors of massless Dirac
fermions. This theory is strongly coupled in the infrared and is expected
to flow, at least for sufficiently large $N_{f}$, to an interacting
conformal field theory (CFT) with an emergent $SU(N_{f})$ flavor symmetry,
at which one observes power-law correlations in order parameters for
several microscopic competing orders~\cite{hermele2004,hermele2005,hermele2008}.

In this story, the first question to be asked concerns the \emph{stability}
of the DSL. Are there relevant operators in this CFT
with the same microscopic lattice symmetries as the DSL? Fermion bilinears
are of course relevant, but always violate microscopic symmetries~\cite{hermele2005,hermele2008,song2019a,song2020}. Of special concern
are monopole operators in CQED$_{3}$~\cite{polyakov1975,polyakov1977,polyakov2021,borokhov2002},
which have their origin in the compactness of the emergent gauge field
that results from the parton construction on the lattice. At least
for sufficiently large $N_{f}$, all monopole operators are irrelevant~\cite{hermele2004,grover2014,borokhov2002,pufu2014} and CQED$_{3}$
remains in a deconfined phase, thus guaranteeing stability of the
DSL. In contrast, the fate of the DSL for small $N_{f}$, including
the value of interest $N_{f}\!=\!4$, is murkier. The issue is the
possible renewed relevancy of monopoles, in which case one then has
to determine if there are monopoles with the same symmetries as the
microscopic realization of the DSL on a given lattice. Correctly determining
how monopole operators transform under lattice symmetries (i.e., their
``quantum numbers'') has been the subject of a longstanding theoretical program~\cite{alicea2005,alicea2005b,alicea2006,alicea2008,hermele2008,song2019a,song2020}.
To be specific, as monopole operators in CQED$_{3}$ are dressed by
fermion zero modes~\cite{borokhov2002,jackiw1984,shankar2021}, their
transformation under lattice symmetries has two contributions: from
the zero modes themselves, and from a $U(1)_{\mathrm{top}}$ phase shift
of the bare monopole interpreted as a Berry phase obtained on dragging
the monopole through a Dirac sea. (Here $U(1)_{\mathrm{top}}$ denotes the $U(1)$ topological symmetry of planar $U(1)$ gauge theories, whose global charge is the total magnetic flux.) The latter Berry phase has been difficult to
compute, and a general framework to do so has only recently emerged
in two works by Song \emph{et al.}~\cite{song2019a,song2020}. Their conclusions
indicate that, in realizations of the DSL on bipartite lattices, there
always exist monopoles that transform trivially under all lattice
symmetries of the state. The relevancy of such monopoles will then
destabilize the DSL, and a transition into one of the proximate competing
orders is then expected.

The second part of the DSL story is then determining the various competing
orders for a given microscopic realization of a DSL~\cite{hermele2005,hermele2008,alicea2008,song2019a,song2020,seifert2023,nambiar2023}.
The immediately available ``order parameters'' in the continuum field theory
are the gauge-invariant fermion bilinears $\bar{\psi}t^{a}\psi$, where $\psi$ is
a spinor in the fundamental representation of the $SU(4)$ flavor symmetry group and $t^{a}\!\in\!\mathfrak{su}(4)$.
However, the spontaneous generation of an expectation value for such a fermion bilinear
is not enough to drive the DSL into the corresponding ordered phase,
 for the fermions are still deconfined. To obtain phases with conventional long-range order, one further requires a mechanism
by which the gauge charges confine. This is assumed to be due to monopole
proliferation in the gauge theory, whose consideration we are again
led to. The state-operator correspondence allows one to classify all
monopole operators by their scaling dimension~\cite{borokhov2002,metlitski2008,pufu2013,dyer2013,pufu2014,dyer2015,chester2016,chester2018,boyack2018,dupuis2019,dupuis2021,dupuis2022,chester2022}. Combined with the methods
developed in Refs.~\cite{song2019a,song2020} to compute the quantum
numbers of the monopoles, one can determine the correct monopoles
to add to the Lagrangian. As argued in those references,\textbf{ }the
transition from the DSL into a proximate conventionally ordered phase
then consists of a two-step process in which a fermion bilinear is
first spontaneously generated, due for instance to a sufficiently strong symmetry-allowed four-fermion interaction~\cite{xu2008}, followed by the
proliferation of the relevant monopoles to drive confinement. In certain cases, the fermion bilinear does not encode all the broken symmetries of a given microscopic order, and monopole proliferation is responsible for breaking the remaining symmetries.

To construct these monopole operators, Ref.~\cite{borokhov2002}
utilized the conformal invariance of massless CQED$_{3}$ at large
$N_{f}$ and defined monopole operators as states in the large-$N_f$ CFT in a background flux on $S^{2}\!\times\!\mathbb{R}$. In
this paper, we use the definition of monopole operators as instanton
defects in the path integral~\cite{polyakov1977,polyakov2021,polyakov1975,murthy1990,unsal2009}
to explicitly reconstruct these directly on $\mathbb{R}^{3}$ as terms
in an effective Lagrangian, in the specific context of a DSL. Moreover,
our construction is not reliant on conformal symmetry. 
Indeed, we
specifically focus on the dynamics of confinement once a fermion mass
$\bar{\psi}t^{a}\psi$ is added to the DSL Lagrangian. We find that
such an ``adjoint mass'' results in the existence of Euclidean zero
modes (of the 3D massive Dirac operator) bound to instantons, distinct from the
zero-\emph{energy} modes that appear in the massless limit. Resumming
the instanton gas results in the generation of an instanton-induced
term in the effective Lagrangian dubbed the 't Hooft vertex~\cite{shankar2021,shankar2022,affleck1982,thooft1976,thooft1976a,thooft1986},
which in this case turns out to be equivalent to the zero mode-dressed monopole
operator found in the CFT approach. For ordered phases with (broken) symmetries
fully captured by a fermion mass, we show that requiring the associated
't Hooft vertex to satisfy the same symmetries can be sufficient to
compute monopole quantum numbers under microscopic symmetries. As observed
in Refs.~\cite{song2019a,song2020}, the DSL on square and honeycomb
lattices possesses such proximate orders, in contrast to non-bipartite
lattices. 

The rest of the paper is structured as follows. After a review of
the parton construction of the DSL in Sec.~\ref{sec:revDSL}, we
organize the effects of monopoles in the path integral as an instanton-gas sum in Sec.~\ref{subsec:EZM}, where it is also shown that such
instanton-bound zero modes cause the path integral to vanish. The physical meaning
of these Euclidean zero modes, and their relation to zero-energy
modes found in previous constructions in the literature, are discussed
in Sec.~\ref{subsec:ezmH}. Section~\ref{subsec:resum} discusses the
technical computation of the 't Hooft vertex by resumming the instanton
gas. This 't Hooft vertex is rewritten by introducing ``zero-mode
operators'' in Sec.~\ref{sec:zmmon}, which reveals the relation
to monopole operators constructed in the CFT approach. After discussing
the continuum symmetries of the instanton-induced monopole operators, we comment in Sec.~\ref{sec:mqnum} on
their quantum numbers under lattice symmetries for bipartite lattices, and finally conclude in Sec.~\ref{sec:concl}.

\section{Review of Dirac spin liquids\label{sec:revDSL}}

For concreteness, we consider the spin-1/2 antiferromagnetic Heisenberg
model,
\begin{equation}
H=\sum_{ij}\c{J}_{ij}\bm{S}_{i}\cdot\bm{S}_{j},
\end{equation}
on an arbitrary planar lattice, although one really has in mind an
equivalence class of lattice models differing by symmetry-allowed
terms. To obtain spin-liquid states, one typically begins with a parton
representation~\cite{wen2007},
\begin{equation}
\bm{S}_{i}=\frac{1}{2}\sum_{\alpha,\beta=\uparrow,\downarrow}c_{i\alpha}^{\dagger}\bm{\sigma}_{\alpha\beta}c_{i\beta},\label{eq:pdecom}
\end{equation}
where $c_{i\alpha}$ are fermions of spin-1/2 and $\bm{\sigma}\!=\!(\sigma_{x},\sigma_{y},\sigma_{z})$
is the Pauli vector. Since the local spin-1/2 Hilbert space is only
two-dimensional, the parton representation introduces a gauge redundancy,
and one must project out unphysical states using the single-occupancy
constraint:
\begin{equation}
\sum_{\alpha}c_{i\alpha}^{\dagger}c_{i\alpha}=1.\label{eq:UVconstraint}
\end{equation}
The gauge group can be seen to be $SU(2)$, for a local $SU(2)$ rotation
of the Nambu spinor $(c_{i\uparrow}\quad c_{i\downarrow}^{\dagger})$
leaves the spin operator (\ref{eq:pdecom}) invariant.

The Heisenberg model then becomes a quartic interaction of fermions,
which can be exactly decoupled inside a path integral using Hubbard-Stratonovich
(HS) fields, as a prelude to mean-field theory. Motivated by a search
for translationally and rotationally invariant spin liquids\footnote{While breaking spin-rotation symmetry does not preclude a spin-liquid ground
state~\cite{takagi2019}, the Dirac spin liquid is a state that
preserves this symmetry.}, the most general decoupling consistent with these requirements results
in a Lagrangian (assuming sums over repeated spin indices):
\begin{align}
L=\sum_{i}c_{i\alpha}^{\dagger}\partial_{\tau}c_{i\alpha} & -\sum_{ij}\frac{\c{J}_{ij}}{4}(c_{i\alpha}^{\dagger}z_{ij}c_{j\alpha}\!+\!\mathrm{h.c.})
  -\sum_{ij}\frac{\c{J}_{ij}}{4}(\epsilon_{\alpha\beta}c_{i\alpha}^{\dagger}w_{ij}c_{j\beta}^{\dagger}\!+\!\mathrm{h.c.})\nonumber \\
 & +\sum_{ij}\frac{\c{J}_{ij}}{4}(\abs{z_{ij}}^{2}+\abs{w_{ij}}^{2})
  -ia_{0}(c_{i\alpha}^{\dagger}c_{i\alpha}\!-\!1).
\end{align}
Here, $a_{0}(\tau)$ is a Lagrange multiplier field that imposes the
half-filling constraint, and $z_{ij}$ and $w_{ij}$ are complex-valued
HS link fields. The saddles of $z_{ij}$ and $w_{ij}$ are respectively
at $c_{i\alpha}^{\dagger}c_{j\alpha}$ and $\epsilon_{\alpha\beta}c_{i\alpha}^{\dagger}c_{j\beta}^{\dagger}$,
so mean-field ans\"atze for $z_{ij}$ and $w_{ij}$ are equivalent to
condensing those fermion bilinears. Introducing the Nambu variables,
\begin{equation}
\psi_{i}=\begin{pmatrix}c_{i\uparrow}\\
c_{i\downarrow}^{\dagger}
\end{pmatrix},\qquad T_{ij}=\begin{pmatrix}z_{ij} & w_{ij}\\
w_{ij}^{\dagger} & -z_{ij}^{\dagger}
\end{pmatrix},\label{eq:nambu}
\end{equation}
and Pauli matrices $\tau^l$, $l\!=\!1,2,3$ that act in this Nambu space, and relabeling $a_0\!\rightarrow\!a_0^3$, the Lagrangian can be rewritten as:
\begin{equation}
L=\sum_{i}\psi_{i}^{\dagger}(\partial_{\tau}-ia_{0}^{l}\tau^{l})\psi_{i}-\sum_{ij}\frac{\c{J}_{ij}}{4}(\psi_{i}^{\dagger}T_{ij}\psi_{j}+\mathrm{h.c.})
+\sum_{ij}\frac{\c{J}_{ij}}{8}\tr T_{ij}^{\dagger}T_{ij},\label{eq:Lsu2}
\end{equation}
where the half-filling constraint is redundantly imposed using two
more Lagrange multipliers, $a_{0}^{1}$ and $a_{0}^{2}$, to produce
the temporal component $a_{0}\!\equiv\!a_0^l\tau^l$ of an $\mathfrak{su}(2)$ gauge field.
Indeed, the Lagrangian is now invariant under an $SU(2)$ gauge
transformation:
\begin{align}
a_{0}(i) & \to\Omega_{i}(a_{0}+i\partial_{\tau})\Omega_{i}^{\dagger},\nonumber \\
T_{ij} & \to\Omega_{i}T_{ij}\Omega_{j}^{\dagger},\nonumber \\
\psi_{i} & \to\Omega_{i}\psi_{i}.
\end{align}
The Lagrangian (\ref{eq:Lsu2}) is an exact representation of the
spin-1/2 Heisenberg model on an arbitrary lattice, and describes a
lattice $SU(2)$ gauge theory at infinite gauge coupling (i.e., with
no dynamics for the gauge fields), but with the group elements $U_{ij}$
on every link being arbitrary complex matrices instead of $SU(2)$
matrices. However, any complex matrix admits a polar decomposition\textbf{
}
\begin{equation}
T=\sqrt{T^{\dagger}T}U\equiv\rho U,\label{eq:polar}
\end{equation}
where $U$ is unitary, and $\rho$ is positive semi-definite and Hermitian.

At this point, one chooses a mean-field ansatz $\ev{T_{jk}}$ that
renders the parton Hamiltonian quadratic. As $T_{jk}$ is gauge-covariant,
this ansatz generically violates gauge invariance, and the mean-field
Hamiltonian $H_{\mathrm{mf}}$ will not commute with the constraint
$\tr\psi_{i}^{\dagger}\tau^{l}\psi_{i}$. However, some measure of
gauge invariance is restored by considering fluctuations in $T_{jk}$
about its mean-field value. Of these, there are ``amplitude fluctuations''
in $\rho$ and ``phase fluctuations'' in $U$, as evident from (\ref{eq:polar}).
Since $\rho$ only modulates the magnitude of the hopping,
it is expected that the fluctuations of qualitative importance are
those of the ``phase matrix'' $U$. Since we are interested in the
infrared fate of the system, these gauge fluctuations will have dynamics
due to a renormalization of the gauge coupling to finite values under
RG flow of (\ref{eq:Lsu2}). This means the hard gauge constraint
(\ref{eq:UVconstraint}) will be softened in the infrared to 
\begin{equation}
(\partial E^{l})_{i}=\tr\psi_{i}^{\dagger}\tau^{l}\psi_{i},
\end{equation}
where the left-hand side is the lattice divergence of the electric field. It
is understood that the fermions on the right-hand side are now renormalized fermions,
and thus need not obey the hard constraint of the ultraviolet partons originally
used in the parton construction. The mean-field Hamiltonian is then
understood as written in terms of these renormalized partons, dubbed spinons.

Then writing $T_{ij}\!=\!\bar{T}_{ij}\exp(ia_{ij})$ to allow for phase fluctuations,
it is intuitive from (\ref{eq:Lsu2}) that a generic mean-field value $\bar{T}$,
which translates to condensing bilinears of type $c_{i\alpha}c_{j\alpha}^{\dagger}$
and $c_{i\uparrow}c_{j\downarrow}$, might Higgs the
$\mathfrak{su}(2)$ gauge bosons down to some subgroup. A criterion
given by Wen determines the infrared gauge group~\cite{wen2002,wen2007,mudry1994}.
Considering all based loops on the lattice, a collinear flux (in some
direction in $SU(2)$ space) of the mean-field $\bar{T}$ through
all such loops results in a Higgsing of $SU(2)\!\to\!U(1)$, and generic
non-collinear fluxes will break it down to $\mathbb{Z}_{2}$, completely
gapping out all gauge bosons. In contrast, a trivial $SU(2)$
flux $(\propto\!\mathbb{I})$ ensures all the $\mathfrak{su}(2)$
gauge bosons remain massless. We shall be specifically interested
in mean-field states that Higgs $SU(2)\!\to\!U(1)$ on various lattices. Examples
include the staggered flux state on the square lattice~\cite{hermele2005},
or the $\pi$ flux state on the kagome lattice~\cite{hastings2000,ran2007,hermele2008}.
The spinons $(c_{\uparrow},c_{\downarrow})$ in these states have
relativistic dispersions, with generically two Dirac nodes $(\alpha\!=\!\pm)$
in the bandstructure. A linearized description at these nodes with low-energy fermions $\psi_{\alpha\sigma}$, that also accounts for U(1) gauge fluctuations with an emergent gauge field  $a_\mu$, is then given by the continuum (Euclidean) Lagrangian:
\begin{equation}
\mathcal{L}=\bar{\psi}(\slashed{\partial}-i\slashed{a})\psi+\frac{1}{4e^{2}}f^{2},
\end{equation}
where $f_{\mu\nu}\!=\!\partial_\mu a_\nu-\partial_\nu a_\mu$ is the field strength tensor, $\psi$ is a Dirac 2-spinor in the fundamental representation of $SU(4)$, the
gamma matrices $(\gamma_{1},\gamma_{2},\gamma_{3})\!=\!(\gamma_{x},\gamma_{y},\gamma_{z})$
are chosen as the three Pauli matrices, and the Dirac adjoint is $\bar{\psi}\!=\!\psi^{\dagger}\gamma_{3}$.
Since the gauge coupling $e^{2}$ has dimensions of inverse length,
the Lagrangian is expected to be strongly coupled in the infrared, flowing
to an interacting conformal fixed point which we shall call the DSL
fixed point. In the $1/N_f$ expansion, one can show that
this fixed point becomes nearly free, characterized by $e_{\ast}^{2}\!\propto\!N_{f}^{-1}$,
so that in the limit $N_{f}\!\to\!\infty,$ gauge fluctuations are
suppressed and spinons are free~\cite{kim1999,rantner2002,hermele2004}. While it is unclear if this fixed point
persists as $N_{f}$ is lowered to the physically relevant value $N_{f}\!=\!4$,
conformal invariance at large $N_{f}$ provides an accessible window
to find relevant operators that can destabilize the DSL. Of central
importance are monopole operators arising from the compactness of
$a$, which when proliferated act to confine spinons into gauge-neutral
spins~\cite{polyakov1975,polyakov1977,polyakov2021}, yielding conventional
phases of the parent spin system.

These monopole operators can be defined at the large-$N_f$ DSL fixed point via the state-operator correspondence in radial quantization,
by considering free fermions on a sphere containing a monopole (plus fluctuations controlled by the $1/N_f$ expansion)~\cite{borokhov2002}.
The monopole with smallest scaling dimension corresponds to the ground
state of the fermions. In a $2\pi$ flux background created
by a minimal monopole, there is one zero-energy mode per flavor of
relativistic fermion as required by the Atiyah--Singer index theorem. To obtain a gauge-invariant state respecting
the constraint (\ref{eq:UVconstraint}), half of the four zero-energy
modes have to be filled. There are thus $\textstyle{\begin{psmallmatrix}
    4 \\ 2
\end{psmallmatrix}=6}$ monopole operators of
minimal charge. If there is a symmetry-allowed relevant monopole,
then the DSL is an unstable critical point separating ordered phases.
If all monopoles are irrelevant, then there is no confinement and
a stable DSL is obtained. However, there could be other interactions
that drive symmetry-breaking by generating a fermion mass, allowing
a previously disallowed monopole to then condense, causing confinement.
We will now proceed to explicitly construct these monopole operators
without relying on conformal invariance. As a byproduct of such a
construction, we will obtain the exact monopole that proliferates
for a given pattern of symmetry breaking described by the ``adjoint
masses'':
\begin{equation}
M^{a}=m\bar{\psi}t^{a}\psi,\qquad t^{a}\in\{\sigma^{i},\mu^{i},\sigma^{i}\mu^{j}\},
\end{equation}
where $\sigma_{i},\mu_{i}$ are Pauli matrices that act on spin and
nodal indices respectively. The 15 mass terms considered above are the most general (Hermitian and Lorentz-invariant) fermion masses that do not radiatively generate a Chern-Simons term for the gauge field $a_\mu$. Indeed, along with the identity matrix, they form a basis for the space of all $4\!\times\!4$ Hermitian matrices. The identity itself, that is the fermion mass $\bar{\psi}\psi$, generates a Chern-Simons term for $a_\mu$ and is not expected to lead to a symmetry-broken phase with confined gauge charges. In the CFT picture, an adjoint mass spoils
conformal invariance and splits the degeneracy between the four zero-energy modes, causing one particular combination of the six monopole
operators to lower its scaling dimension compared to the rest~\cite{nambiar2023}.
Our construction will directly yield this monopole, and by varying
the adjoint mass yields all linearly independent monopole operators.

\section{The 't Hooft vertex}

The basic idea behind our construction is to (1) formulate the instanton problem in its original Euclidean path-integral language, rather than the canonical-quantization formalism of CFT, and (2) utilize semiclassical instanton calculus~\cite{thooft1976,thooft1976a,thooft1986,unsal2009} to resum a monopole-instanton
gas in the presence of massive fermions~\cite{shankar2021}. We show
that the existence of instanton-bound fermion zero modes (ZMs) of
the \emph{Euclidean} Dirac operator on $\mathbb{R}^3$ cause transition amplitudes to vanish
unless fermion insertions can ``soak up'' these ZMs in the path-integral measure. This is in contrast to the case of massless Dirac fermions, for which no normalizable Euclidean ZMs exist~\cite{marston1990a}. These insertions
will then ``dress'' the bare monopole operator that simply creates
$2\pi$ flux in the gauge theory. 

\subsection{Euclidean fermion zero modes\label{subsec:EZM}}

To set up our semiclassical calculation, we decompose the emergent gauge field as:
\begin{equation}
a=\mathcal{A}+\delta a,
\end{equation}
where $\mathcal{A}$ is a monopole-instanton solving the Euclidean
equations of motion, and $\delta a$ describes smooth fluctuations
(photons) around the instanton solution. Temporarily neglecting the
coupling of fermions to photons\footnote{This is justified in a large-$N_{f}$ approximation, but one can improve
the calculation by considering fluctuations around the instanton just
as in Ref.~\cite{pufu2014}.}, the partition function can be written as a sum over an instanton
gas~\cite{shankar2021}:
\begin{multline}
Z\!=\! \int\! Da\,e^{-\frac{e^{2}}{2}\int d^{3}x\,(\partial_{\mu}\sigma)^{2}}\sum_{N=0}^{\infty}\frac{1}{N!}
 \prod_{k=1}^{N}\left(\int d^{3}z_{k}\sum_{q_{k}\in\mathbb{Z}}e^{-q_{k}^{2}/e^{2}\ell}e^{iq_{k}\sigma(z_{k})}
 \int D(\bar{\psi},\psi)e^{-S_{f}[\mathcal{A}(q_{k})]}\right),\label{eq:zsector}
\end{multline}
where $\sigma$ is the dual photon~\cite{polyakov2021}, $N$ is
the number of monopoles in the gas, $q_{k}$ their charges, $z_{k}$
their locations (a collective coordinate), and $q_{k}^{2}/e^{2}\ell$
with short-distance cutoff $\ell$ (on the order of the lattice constant) the action cost for a charge-$q_{k}$
monopole. Finally, $S_{f}[\mathcal{A}(q_{k})]$ is the fermion action
in a \emph{single-instanton} \emph{background} specified by $(q_{k},z_{k})$.
A dilute-gas approximation has been made in the partition above, which
allows one to partition an $N$-instanton background as $\mathcal{A}\!=\!\sum_{k=1}^{N}\mathcal{A}_{(k)},$
describing $N$ well-separated boxes containing a single instanton
each. Assuming a dilute gas of monopoles allows one to bring the fermion
path integral inside the product in (\ref{eq:zsector}), and consider
fermions moving in a single-instanton background instead of that of
a correlated instanton liquid. This is formally accomplished by decomposing
$\psi$ into fields localized in large boxes around each instanton~\cite{thooft1976a}, with zero overlap between boxes. This is justified
in hindsight by the observation that fermion ZMs are exponentially
localized on the instantons.

A brief remark on the validity of the dilute gas approximation is in order. In the pure gauge theory (without matter), monopoles interact via a Coulomb potential, which is known not to form bound states in 3D. The diluteness of this Coulomb gas can be argued using a standard Debye screening theory that assumes weak coupling $e^2\ell\!\ll\!1$, as in Polyakov's original work \cite{polyakov2021,polyakov1977}. In the theory with fermion matter considered in this work, we will see that there exist pair exchanges of fermions between monopoles. Such pair exchanges are expected to correct the Coulomb potential between bare monopoles. However, since the fermions considered are massive, we suspect that such a correction is not enough to overwhelm the long-range Coulomb interaction and result in monopole bound states that would invalidate a dilute gas approximation.

This paper will only be concerned with monopole operators of lowest
charge $q\!=\!\pm1$, although the computation straightforwardly generalizes
to higher charges in an obvious way. The fermion path integral in Eq.~(\ref{eq:zsector}),
which we separately write as:
\begin{equation}
Z_{f}[\mathcal{A}_{q}]=\int D(\bar{\psi},\psi)e^{-\int\bar{\psi}(\slashed{\partial}-i\slashed{\mathcal{A}}_{q}+mt^{a})\psi},
\end{equation}
evaluates to zero for a gauge-field configuration $\mathcal{A}_{q}$ with nonzero monopole charge $q$. This is because
the Euclidean Dirac operator:
\begin{equation}
\mathfrak{D}_{q}=\slashed{\partial}-i\mathcal{A}_{q}+mt^{a},\label{eq:dop}
\end{equation}
has nontrivial ZMs in an instanton background. Unlike
zero-\emph{energy} modes of the Hamiltonian~\cite{jackiw1976} that
are typically bound to solitons, these zero modes of $\mathfrak{D}$
are bound to instantons. The relation between energy ZMs and these
Euclidean ZMs will be further elucidated in Sec.~\ref{subsec:ezmH}.

Explicit solutions for these ZMs are obtained in Appendix~\ref{app:zmdrop}.
For a fixed mass $mt^{a}\!\in\!\mathfrak{su}(4)$ with $m\!>\!0$,
the normalizable ZMs of $\mathfrak{D_{\pm}}$ in $q\!=\!\pm1$ backgrounds
are (respectively):
\begin{align}
u_{+}^{(i)}(r,\theta,\varphi)&\!=\!  \frac{\sqrt{2m}}{r}e^{-mr}\mathcal{Y}_{1,0,0}^{1/2}(\theta,\varphi)\ket{i}_{a},\quad i\!=\!2,4;\label{eq:zm+}\\
u_{-}^{(i)}(r,\theta,\varphi)&\!=\!  \frac{\sqrt{2m}}{r}e^{-mr}\mathcal{Y}_{-1,0,0}^{1/2}(\theta,\varphi)\ket{i}_{a},\quad i\!=\!1,3,\label{eq:zm-}
\end{align}
and those of $\mathfrak{D}_{\pm}^{\dagger}$ are respectively:
\begin{align}
v_{+}^{(i)}(r,\theta,\varphi)&\!=\!  \frac{\sqrt{2m}}{r}e^{-mr}\mathcal{Y}_{1,0,0}^{1/2}(\theta,\varphi)\ket{i}_{a},\quad i\!=\!1,3;\label{eq:zm+dag}\\
v_{-}^{(i)}(r,\theta,\varphi)&\!=\!  \frac{\sqrt{2m}}{r}e^{-mr}\mathcal{Y}_{-1,0,0}^{1/2}(\theta,\varphi)\ket{i}_{a},\quad i\!=\!2,4,\label{eq:zm-dag}
\end{align}
where the four eigenvectors of the $\mathfrak{su}(4)$ mass are defined by:
\begin{align}\label{taeigenvec}
t^{a}\ket{i}_{a}\!=\!(-1)^{i}\ket{i}_{a},\hspace{5mm}i=1,2,3,4,
\end{align}
and $\mathcal{\mathcal{Y}}_{q,j,M}^{j\pm1/2}$ are monopole spinor
harmonics as defined in Appendix~\ref{app:zmdrop}. As discussed in
the subsequent section, gauge invariance mandates that only two of
these ZMs can be filled in any fixed instanton background. It will
turn out to be sufficient to consider the ZMs $u_{+}^{(i)}$ and $v_{-}^{(i)}$
of $\mathfrak{D}_{+}$ and $\mathfrak{D}_{-}^{\dagger}$ to obtain
nearly all the results in this paper. As shown in Sec.~\ref{subsec:resum},
these lead to spontaneous fermion pair creation (in an instanton background)
and annihilation events (in an anti-instanton background).

The topological guarantee of these Euclidean ZMs is provided by their
relation to \emph{energy ZMs} of a\emph{ massless} Dirac Hamiltonian
in a static $2\pi q$ flux background, which are protected by an Atiyah--Singer
index theorem~\cite{jackiw1984}. Indeed, the Euclidean ZMs above
in the limit $m\!\to\!0$ (ignoring the normalization) have precisely
the form of the energy ZMs observed in radial quantization on $S^{2}\times\!\mathbb{R}$,
on recognizing the Weyl rescaling factor $r^{-1}\!=\!\exp(-\tau)$~\cite{borokhov2002}. A nonzero mass gaps out these energy ZMs,
which reincarnate as normalizable (exponentially localized) ZMs of the Euclidean Dirac operator $\mathfrak{D}$.
As we show below, the physical consequence of these Euclidean ZMs
is that instanton events are correlated with fermion-number violating
processes.

\subsection{Euclidean fermion zero modes in a Hamiltonian view\label{subsec:ezmH}}

We first present an intuitive argument for the heretofore claimed
fermion-number violating processes caused by instantons. Instead of
modeling the instanton as a point source
of flux spatially localized in 2D, we can distribute the $2\pi q$ flux uniformly across the
area $A$ of a finite system. This is physically reasonable as a nonzero gauge
coupling will supply monopoles with momentum, effectively delocalizing
them. What is important is that the total flux through the system
can only jump discretely through instanton events. Massive\emph{ }Dirac
fermions under this uniform magnetic field $2\pi q/A$ are then housed
in relativistic Landau levels (for each flavor),
\begin{align}
E_{n\pm} & =\pm\sqrt{2\pi n\abs{q}A^{-1}+m^{2}},\qquad n\geq1,\nonumber \\
E_{0} & =m\sgn(q),
\end{align}
where $E_0$ is the ``zero Landau level'' obtained in the massless
limit. The degeneracy of the levels is $\abs{q}$, so that for $q\!=\!+1$
there is precisely one ``zero mode'' per flavor of Dirac fermion, in agreement with the Atiyah--Singer index theorem,
giving a total of four modes for $N_{f}\!=\!4$.

\begin{figure}[h]
\begin{centering}
\includegraphics[width=1\textwidth,height=0.6\textwidth,keepaspectratio]{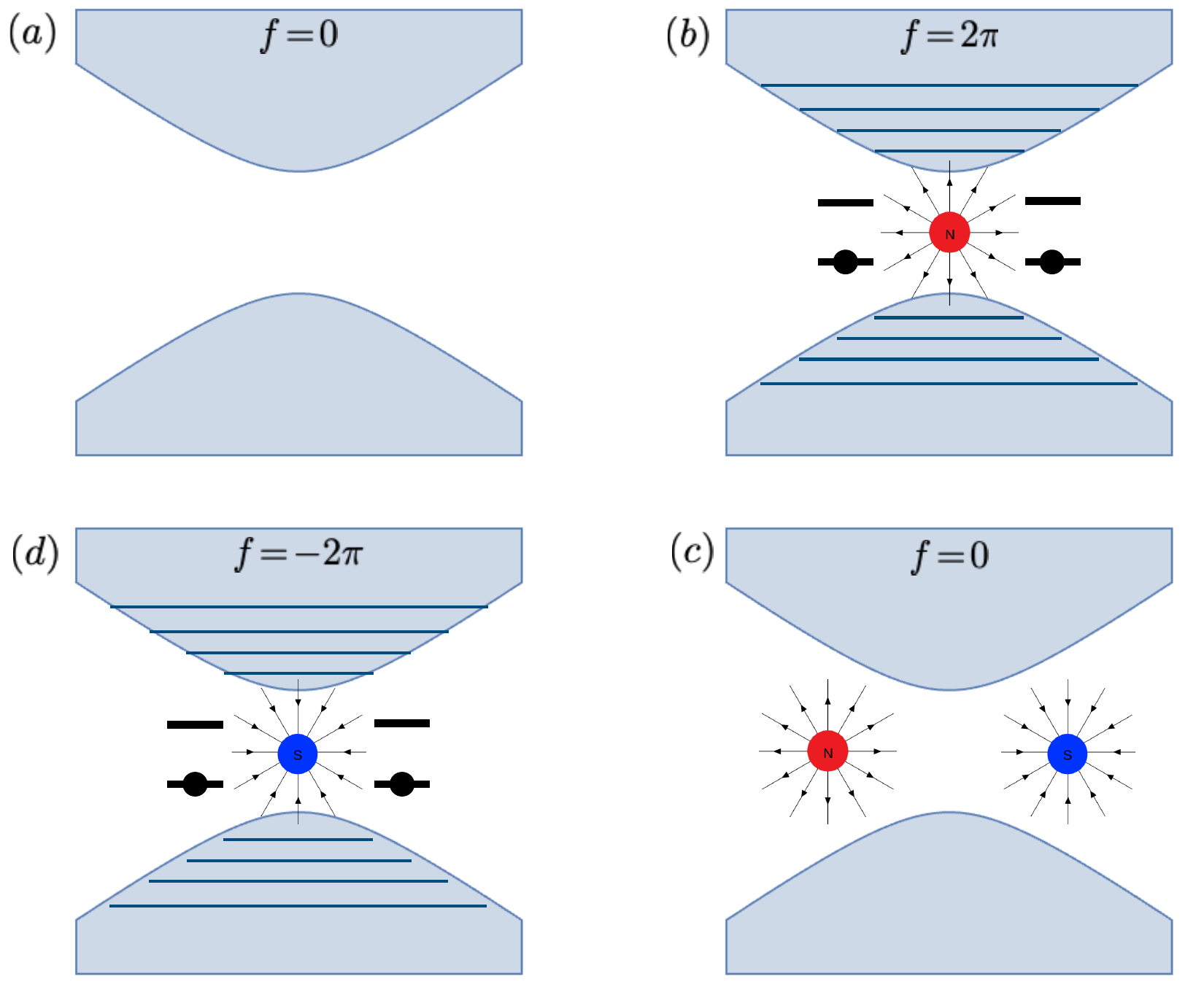}
\par\end{centering}
\caption{Massive Dirac fermions in various flux backgrounds at half-filling. Clockwise from top left: [(a) to (b)] Instanton $f\!=\!0\to\!2\pi$ accompanied by spinon pair creation in two midgap modes. [(b) to (c)] Anti-instanton $f\!=\!2\pi\to\!0$ accompanied by pair annihilation of spinons in occupied midgap modes. [(c) to (d)] Anti-instanton $f\!=\!0\to\!-2\pi$ with pair creation of spinons in two midgap modes. [(d) to (a)] Instanton $f\!=\!-2\pi\to\!0$ accompanied by pair annihilation of spinons in the occupied midgap modes.}
\label{fig:instzms}
\end{figure}

In preparation for an interpretation of instanton events, let us imagine
adiabatically dialing the flux from 0 to $2\pi$. In the zero-flux
limit, we simply have two bands formed by gapping a Dirac cone [Fig.~\ref{fig:instzms}(a)]. The
single-occupancy constraint (\ref{eq:UVconstraint}) ordained by the parton decomposition (\ref{eq:pdecom}), which is equivalent to a Gauss' law constraint, mandates a half-filling of these bands (for each flavor of Dirac fermion). 
As a $2\pi$ flux is adiabatically turned on, the $E_{n\pm}$ levels
evolve in perfect tandem out of the upper and lower bands, while the
``zero mode'' captures the \emph{spectral asymmetry} of the Hamiltonian.
Depending on the relative sign of $m$ and $q$, it either descends
from the upper band {[}$\sgn(mq)\!>\!0${]} or ascends from the lower
one $[\sgn(mq)\!<\!0]$. Since we are working with $\mathfrak{su}(4)$-valued
masses $mt^{a}$ that preserve time-reversal (TR) invariance, it follows
that there are a total of 4 displaced energy ``ZMs'', two with energy
$m$ and two with energy $-m$ [Fig.~\ref{fig:instzms}(b)]. Gauge invariance (i.e., the single-occupancy constraint) again requires us
to fill two of these modes. The ground state is \emph{uniquely }obtained
by filling the two negative-energy modes. This is to be contrasted
with the massless limit, in which all four modes are degenerate at
zero energy, and there are six possible ways to fill two of them.
Selecting a specific $\mathfrak{su}(4)$-mass $mt^{a}$ gaps the four
degenerate ZMs in a TR-invariant manner, selecting precisely two of
them to fill.

Therefore, when the flux tunnels from $0\!\to\!2\pi$ by means of an instanton, two negative-energy
modes suddenly appear in the spectrum. If these remain unfilled\footnote{It is assumed that we are at sufficiently low temperature that the
leading order contribution is the filling of the two \emph{negative
}modes rather than the positive mid-gap modes that are also present.
In any case, we shall see in the next section that the selection of
two modes automatically falls out of the calculation.}, the instanton would have caused an unphysical transition from a
gauge invariant state to a non-invariant state violating the half-filling
condition. The resolution is that an instanton event \emph{must }be
accompanied by fermion pair creation in the two new unfilled levels.
Proceeding then in reverse from $2\pi\!\to\!0$ flux sectors by means
of an ``anti-instanton'', we immediately observe that anti-instantons
should cause fermion pair annihilation as the two ``ZMs'' disappear
into the lower bands [Fig.~\ref{fig:instzms}(c)].

These considerations lead to the conclusion that instantons cause
fermion pair creation and annihilation. However, such processes must
be reflected in an appropriate effective Lagrangian by means of ``dressed''
monopole operators of the form:
\begin{equation}
\mathcal{M}\bar{\psi}\Delta_{+}\bar{\psi}^{\intercal}+\mathcal{M^{\dagger}}\psi^{\intercal}\Delta_{+}^\dagger\psi,\label{eq:Manz}
\end{equation}
where $\mathcal{M}$ is a ``bare'' monopole operator that creates
$2\pi$ flux, $^{\intercal}$ denotes the transpose, and $\Delta_{+}$ is a vertex factor valued in $\mathfrak{su}(4)$
that will select precisely two flavors from the four $\psi_{\alpha\sigma}$
to fill the two displaced energy ZMs just discussed. Determination
of this vertex factor for a specific $\mathfrak{su}(4)$ mass is
one of the central goals of this work, a task that shall be taken
up in the next section.

Finally, the above considerations can be equally applied to flux tunneling
from $0\!\to\!(-2\pi)$, which leads to the conclusion that anti-instantons
can also create fermions [Fig.~\ref{fig:instzms}(d)]. This would yield a vertex contribution
\begin{equation}
\mathcal{M^\dagger}\bar{\psi}\Delta_{-}\bar{\psi}^{\intercal}+\mathcal{M}\psi^{\intercal}\Delta_{-}^\dagger\psi.\label{eq:Manz2}
\end{equation}

\subsection{Resummation of the instanton gas\label{subsec:resum}}

In this section, the intuitive picture sketched in the previous section
will be formally laid out in the path-integral framework, and the
monopole operators (\ref{eq:Manz}-\ref{eq:Manz2}) completely determined by a resummation
of the instanton gas in the partition function (\ref{eq:zsector}).
To do so in the presence of ZMs of the Euclidean Dirac operator, we shall use
a slight variant of the technique originally devised by 't Hooft in
his resolution of the $U(1)$ problem in $\mathrm{QCD}_{4}$~\cite{thooft1976,thooft1976a,thooft1986}.
More technical details can be found in Ref.~\cite{shankar2021};
see also Ref.~\cite{affleck1982} for a symmetry-based argument.
An analogous calculation for $SO(N)$ gauge theory with Majorana matter
was done in Ref.~\cite{shankar2022}, which studied confinement
transitions out of a chiral spin liquid. Readers uninterested
in the technical details of the calculation can safely proceed to
the next section with just the final result [Eq.~(\ref{eq:vertex})] in hand.

As observed in Sec.~\ref{subsec:EZM}, the fermion path integral
vanishes in a nontrivial instanton background, implying that only
the sector with zero instanton charge contributes to the partition
function itself in Eq.~(\ref{eq:zsector}). However, sectors
with nonzero charge will contribute to correlation functions that
can ``soak up'' the ZMs (to be explained below). From the discussion
in the previous section, we expect these to be correlators of the
form $\ev{\psi\psi}$. This is best seen with mode expansions of the
spinons $(\psi,\bar{\psi)}$ in eigenfunctions of the self-adjoint
operators $(\mathfrak{D}_{+}^{\dagger}\mathfrak{D}_{+},\mathfrak{D_{+}\mathfrak{D}}_{+}^{\dagger})$
for a $q\!=\!+1$ background:
\begin{align}
\psi & =u_{2+}(x\!-\!z_{+})\eta_{2}\!+\!u_{4+}(x\!-\!z_{+})\eta_{4}+\!\sideset{}{'}\sum_{i}w_{i}(x\!-\!z_{+})\xi_{i},\nonumber \\
\bar{\psi} & =\sideset{}{'}\sum_{i}\bar{w}_{i}(x\!-\!z_{+})\bar{\xi}_{i},\label{eq:fmodexp}
\end{align}
where $w_{i}$ are nonzero modes (indicated by the primed sums) of
$\mathfrak{D}_{+}^{\dagger}\mathfrak{D}_{+}$, which occur in pairs
with $\bar{w}_{i}$ of $\mathfrak{D}_{+}\mathfrak{D}_{+}^{\dagger}$,
and $\{\eta,\xi,\bar{\xi}\}$ are Grassmann numbers to ensure the correct Fermi
statistics. The functions $u_{i+}(x\!-\!z_{+})$ are the ZMs of $\mathfrak{D}_{+}^{\dagger}\mathfrak{D}_{+}$,
localized on a charge $+1$ instanton at $z_{+}$, whose explicit
expressions are given in Eq.~(\ref{eq:zm+}). Only two ZMs have
been included as mandated by the gauge-invariance arguments in Sec.~\ref{subsec:ezmH},
and the ZMs $v_{+}^{(i)}$ of $\mathfrak{D}_{+}^{\dagger}$ have not
been ``filled'' by including them in the mode expansion of $\bar{\psi}$.
Strictly speaking, we should sum over all possibilities by doing a separate
calculation that only includes the two ZMs of $\mathfrak{D}_{+}^{\dagger}$
and not those of $\mathfrak{D}_{+}$. However, it will be easy to
write down the result of such a calculation after our considerations
below.

The functional measure can now be defined as:
\begin{equation}
D(\bar{\psi},\psi)=d\eta_{2}\,d\eta_{4}\sideset{}{'}\prod_{i}d\bar{\xi}_{i}d\xi{}_{i},
\end{equation}
where the prime again denotes the exclusion of ZMs in the product.
Since the ZMs $\{\eta_{2},\eta_{4}\}$ do not appear in the Lagrangian
$\bar{\psi}\mathfrak{D}_{+}\psi$, the Grassmann integrals over these
cause the partition function to vanish. However, pair correlators of the form
$\ev{\psi\psi}$ involve enough insertions to ``soak up'' the ZMs
in the measure and produce a nonzero path integral. An explicit calculation,
using the mode expansions (\ref{eq:fmodexp}), shows that:
\begin{equation}
\ev{\psi^{a}(x)\psi^{b}(y)^{\intercal}}_{+}=-K_{+}u_{2+}^{[a}(x-z_{+})u_{4+}^{b]}(y-z_{+})^{\intercal},\label{eq:gorkov}
\end{equation}
where $a,b$ are $SU(4)$ indices that have been here antisymmetrized (i.e., $v^{[a}w^{b]}\!\equiv\!v^aw^b\!-\!v^bw^a$), and $K_{+}$ is the fermion path
integral over the nonzero modes $(\xi,\bar{\xi})$ in the instanton
background. While this amplitude looks neither Lorentz nor gauge invariant
at present, we reassure the reader that these issues will be addressed
towards the end of the calculation.

Since the fermion ZMs are exponentially bound to the instanton with a width
$m^{-1}$, this result shows that anomalous correlations also decay exponentially away from the instanton with a length scale $m^{-1}$. This
also reinforces the conclusion reached intuitively in the previous
section; the fermion vacuum in the presence of $2\pi$ flux has two
additional fermions compared to the one with zero flux. A transition between the two states is possible only if these two extra
fermions are annihilated, and this is precisely what the $\psi\psi$ insertion achieves.

We now ask for an effective Lagrangian that reproduces such correlation
functions, which will amount to resumming or ``integrating out''
instantons. In the pure gauge theory, it is well known that the result
of such a resummation is a sine-Gordon term $\propto\cos(\sigma)$ that gaps
out the dual photon $\sigma$ in the infrared~\cite{polyakov1975,polyakov1977,polyakov2021}.
With fermionic matter, a $2\pi$ flux is associated with two additional fermions,
so we expect an effective Lagrangian to contain a term of the form
$e^{i\sigma}\bar{\psi}\Delta_{+}\bar{\psi}^{\intercal}$. To determine $\Delta_{+}$,
let us perturb with a generic anti-symmetrized source\footnote{One should strictly add $\psi^{\intercal} J\psi+\mathrm{h.c.}$, but the conjugate
term cannot soak up the zero modes in the path-integral measure in the $q\!=\!+1$
sector, so we drop it to reduce clutter.} $\psi(x)^{\intercal}J(x,y)\psi(y)$, with suppressed $\mathfrak{su}(4)$ and
Lorentz indices, and perturbatively expand to $\mathcal{O}(J)$:
\begin{align}
Z_{f}[\mathcal{A}_{+},J]&= \int D(\bar{\psi},\psi)e^{-\int\bar{\psi}(\slashed{D}_{+}+mt^{a})\psi-\int\psi^{\intercal} J\psi}\nonumber \\
&= \,K_{+}\int d^{3}x\,d^{3}y\,u_{2+}^{\intercal}(x-z_{+})J(x,y)u_{4+}(y-z_{+}) + \mathcal{O}(J^{2}),\label{eq:JZf}
\end{align}
where the second line is obtained by using the mode expansions (\ref{eq:fmodexp}),
and $K_+$ is the fermion integral over non-ZMs as in Eq.~(\ref{eq:gorkov}).
Our arguments in this and previous sections have indicated that such
an amplitude can be reproduced by a path integral of the form~\cite{shankar2021}:
\begin{multline}
I_{+}[J]=\int D(\bar{\psi},\psi)\,e^{-\int\bar{\psi}(\slashed{\partial}+mt^{a})\psi-\int\psi^{\intercal} J\psi}\\
\times\int d^{3}x'\,d^{3}y'\,C_{+}\,\bar{\psi}(x')\omega_{+}(x'\!-\!z_{+})\zeta_{+}(y'\!-\!z_{+})^{\intercal}\bar{\psi}(y')^{\intercal},
\end{multline}
where the vertex $\Delta_{+}$ has been written as a dyadic product $\omega_{+}\zeta_{+}^{\intercal}$
of vectors with possible spinor and $\mathfrak{su}(4)$ indices. We
will determine $C_{+},\,\omega_{+},\,\zeta_{+}$ by demanding equality
with Eq.~(\ref{eq:JZf}) to $\mathcal{O}(J)$. Note that the fermions are no longer in
a flux background in $I_{+}[J]$. Expanding the above integral to
$\mathcal{O}(J)$ and Wick contracting gives:
\begin{equation}
I_{+}[J]=C_{+}\int d^{3}(x,x',y,y')
[G_{f}(x-x')\zeta_{+}]^{\intercal}J(x,y)[G_{f}(y-y')\omega_{+}],
\end{equation}
where $G_{f}\!=\!\ev{\psi\bar{\psi}}_{0}$ is the free fermion propagator.
Comparing with Eq.~(\ref{eq:JZf}) and demanding equality gives the vertex
factors:
\begin{align}
C_{+} & =K_{+}=\mathrm{det}'\mathfrak{D}_{+},\nonumber \\
\zeta_{+} & =G_{f}^{-1}u_{2+}\approx-2\sqrt{2}\pi\mathcal{Y}_{1,0,0}^{1/2}(\theta,\varphi)\ket{2}_{a},\nonumber \\
\omega_{+} & =G_{f}^{-1}u_{4+}\approx-2\sqrt{2}\pi\mathcal{Y}_{1,0,0}^{1/2}(\theta,\varphi)\ket{4}_{a},\label{eq:+vert}
\end{align}
where we have used explicit expressions for the free propagator (Appendix~\ref{sec:xdirac}) and ZM solutions [Eq.~(\ref{eq:zm+})], and
the approximation holds at distances $r\!\gg\!m^{-1}$ (the width of
the instanton-bound ZM). One can now replace $Z_{f}[\mathcal{A}_{+},J]$
with $I_{+}[J]$ in the instanton-gas sum appearing in the partition function
(\ref{eq:zsector}).

To obtain a path integral $I_{-}[J]\!=\!Z_{f}[\mathcal{A}_{-},J]$
in the anti-instanton sector, the calculation above should be repeated
with ZMs of $\mathfrak{D}_{-}\mathfrak{D}_{-}^{\dagger}$ in the mode
expansion of $\bar{\psi}$.\textbf{ }One can write down the result
based solely on reflection positivity (not reality) of the
Euclidean action~\cite{osterwalder1973,wetterich2011}, but since this is somewhat subtle as we shall see
later, it is more prudent to just repeat the above calculation. The
result is:
\begin{equation}
I_{-}[J]=C_{-}\int d^{3}(x,x',y,y')
[G_{f}^{\dagger}(x-x')\zeta_{-}]^{\intercal}J(x,y)[G_{f}^{\dagger}(y-y')\omega_{-}],
\end{equation}
 with
\begin{align}
C_{-} & =K_{-}=\mathrm{det}'\mathfrak{D}_{-}^{\dagger},\nonumber \\
\zeta_{-} & =(-\sld{\partial}+mt^{a})^{-1}u_{2-}\approx-2\sqrt{2}\pi\mathcal{Y}_{-1,0,0}^{1/2}(\theta,\varphi)\ket{2}_{a},\nonumber \\
\omega_{-} & =(-\sld{\partial}+mt^{a})^{-1}u_{4-}\approx-2\sqrt{2}\pi\mathcal{Y}_{-1,0,0}^{1/2}(\theta,\varphi)\ket{4}_{a}.\label{eq:-vert}
\end{align}
Substituting in $I_{\pm}[J]$ for $Z_{f}[\mathcal{A}_{\pm},J]$ in
the partition function (\ref{eq:zsector}), we obtain:
\begin{multline}
Z[J]=\int D\sigma\,D(\bar{\psi},\psi)\,e^{-S_{0}-\int(\psi^{\intercal} J\psi+\mathrm{h.c.})}
\sum_{N=0}^{\infty}\frac{1}{N!}\prod_{k=1}^{N}\int d^{3}z_{k}\int d^{3}x\,d^{3}y\\
\times\left[-K_{+}e^{i\sigma(z_{k})}\bar{\psi}(x)\mathcal{Y}_{1,0,0}^{1/2}(x\!-\!z_{k})\dyad{2}{4}
\mathcal{Y}_{1,0,0}^{1/2}(y\!-\!z_{k})^{\intercal}\bar{\psi}^{\intercal}(y)+\mathrm{r.c.}\right],\label{eq:ZJf}
\end{multline}
where ``r.c.'' denotes the reflection conjugate\footnote{This is the analog of the hermitian conjugate in Euclidean signature, and is discussed in Sec.~\ref{subsec:ref+}.}, dimensionless constants
have been lumped into $K$, and the free action $S_{0}$ is:
\begin{equation}
S_{0}=\,\int\,d^{3}x\,\left[\frac{e^{2}}{2}(\partial_{\mu}\sigma)^{2}+\bar{\psi}(\slashed{\partial}+mt^{a})\psi\right].
\end{equation}
As remarked below the mode expansions in Eq.~\eqref{eq:fmodexp}, one must also sum over a transition amplitude that involves the two ZMs of $\mathfrak{D}_{+}^\dagger$ but not those of $\mathfrak{D}_{+}$. The calculations leading to Eq.~\eqref{eq:ZJf} clearly indicate that resumming instantons with these ZMs would lead to further insertions of the kind:
\begin{equation}
	-K_{-} e^{-i\sigma(z_k)} \bar{\psi}(x) \c{Y}^{1/2}_{-1,0,0}(x\!-\!z_k)\dyad{1}{3}\c{Y}^{1/2}_{-1,0,0}(y\!-\!z_k)^\intercal \bar{\psi}^\intercal(y)+\mathrm{r.c.},
\end{equation}
where the ZMs \eqref{eq:zm-} and \eqref{eq:zm+dag} have been used. As predicted at the end of Sec.~\ref{subsec:ezmH}, this vertex corresponds to spinon-pair creation by anti-instantons. Including these terms in Eq.~\eqref{eq:ZJf}, re-exponentiating the instanton-gas sum and then setting the source $J$ to zero results in
an instanton-induced contribution to the effective action: the \emph{'t
Hooft vertex},
\begin{align}
S_{\mathrm{inst}}^{a}&= K_{+}\int d^{3}z\,e^{i\sigma(z)}\left[\int d^{3}x\,\bar{\psi}(x)\mathcal{Y}_{1,0,0}^{1/2}(x-z)\right]\dyad{2}{4}
 \left[\int d^{3}y\,\mathcal{Y}_{1,0,0}^{1/2}(y-z)\bar{\psi}(y)\right]^{\intercal}\nonumber \\
&+ K_{-}\int d^{3}z\,e^{-i\sigma(z)}\left[\int d^{3}x\,\mathcal{Y}_{-1,0,0}^{1/2}(x-z)^{\dagger}\psi(x)\right]^{\intercal}
 \dyad{4}{2}\left[\int d^{3}y\,\mathcal{Y}_{-1,0,0}^{1/2}(y-z)^{\dagger}\psi(y)\right]\nonumber\\
&+ K_{-}\int d^{3}z\,e^{-i\sigma(z)}\left[\int d^{3}x\,\bar{\psi}(x)\mathcal{Y}_{-1,0,0}^{1/2}(x-z)\right]\dyad{1}{3}
\left[\int d^{3}y\,\mathcal{Y}_{-1,0,0}^{1/2}(y-z)\bar{\psi}(y)\right]^{\intercal}\nonumber \\
&+ K_{+}\int d^{3}z\,e^{i\sigma(z)}\left[\int d^{3}x\,\mathcal{Y}_{1,0,0}^{1/2}(x-z)^{\dagger}\psi(x)\right]^{\intercal}
 \dyad{3}{1}\left[\int d^{3}y\,\mathcal{Y}_{1,0,0}^{1/2}(y-z)^{\dagger}\psi(y)\right],
 \label{eq:vertex}
\end{align}
where the superscript \emph{a }in $S_{\mathrm{inst}}^{a}$ serves
to remind that this effective interaction is associated with a given
adjoint mass $mt^{a}$, whose eigenvectors $\ket{i}$ feature in the
vertex. However, at this point, we note that the role of the fermion
mass is solely to regulate $K_{\pm}\!=\!(\mathrm{det}'\mathfrak{D}_{\pm}^{\dagger}\mathfrak{D}_{\pm})^{1/2}$ (our discussion below of reflection positivity will imply $K_{+}\!=\!K_{-}\!\equiv\!K$),
and the derived instanton-induced vertex is sensible in the massless limit,
with the functional determinant being regulated in some other way. The adjoint
mass then serves a role similar to a symmetry-breaking source for
a specific ordered state in our calculation. When the massless limit,
which does not ``commute'' with the resummation of the instanton
gas, is taken at the end, the adjoint mass leaves behind in its wake
a monopole which in turn will drive a confining transition into a
proximate ordered state.

\section{Monopole operators and their symmetries\label{sec:zmmon}}

We will now rewrite the 't Hooft vertex~(\ref{eq:vertex}) using ``zero-mode
operators'' in a form that makes explicit its relation to the CFT monopole
operators constructed in Ref.~\cite{borokhov2002}. To this end,
we define the mode operators:
\begin{align}
\bar{c}_{qjM}(z) & =\int d^{3}x\,\bar{\psi}(x)\mathcal{Y}_{qjM}^{j+1/2}(x-z),\nonumber \\
c_{qjM}(z) & =\int d^{3}x\,\mathcal{Y}_{qjM}^{j+1/2}(x-z)^{\dagger}\psi(x),\nonumber \\
c_{\pm1,0,0} & \equiv d_{\pm},\label{eq:zmops}
\end{align}
where flavor indices have been suppressed. These can be thought of
as a spacetime analog of a change of basis with coefficients $\braket{jM}{x}$.
In fact, this follows from a mode expansion of the fermion fields in monopole
harmonics (see Eq.~(7.3) of Ref.~\cite{dyer2013}), and thereby
identifies $c_{qjM}$ as the $\mathbb{R}^{3}$ analog of the ``zero-mode operators'' of Refs.~\cite{borokhov2002,dyer2013}, there
defined in radial quantization on $S^{2}\!\times\!\mathbb{R}$. The
't Hooft vertex (\ref{eq:vertex}) can be written in terms of these
operators as:
\begin{align}
S_{\mathrm{inst}}=K\int d^{3}z\,\Bigl[&e^{i\sigma(z)}\bar{d}_{+}(z)\dyad{2}{4}\bar{d}_{+}(z)^{\intercal}
+e^{-i\sigma(z)}d_{-}(z)^{\intercal}\dyad{4}{2}d_{-}(z)\nonumber\\
&+e^{-i\sigma(z)}\bar{d}_{-}(z)\dyad{1}{3}\bar{d}_{-}(z)^{\intercal}
+e^{i\sigma(z)}d_{+}(z)^{\intercal}\dyad{3}{1}d_{+}(z)\Bigr],\label{eq:finalvert}
\end{align}
which should be understood as the monopole operator spawned by a given
adjoint mass $mt^{a}$ with eigenvectors as in Eq.~(\ref{taeigenvec}). This form makes it manifestly clear that the
$\mathfrak{su}(4)$ part of the vertices, $\dyad{2}{4}$ and $\dyad{1}{3}$, must be antisymmetrized,
in accordance with the observation in Refs.~\cite{borokhov2002,dyer2013}
that monopole operators of minimal charge transform in the antisymmetric
representation of the flavor group with $N_{f}/2$ indices. Before discussing flavor symmetry in greater detail, we first derive how the ZM operators (\ref{eq:zmops}) transform under spacetime
symmetries, reflection positivity, and gauge transformations.

\subsection{Spacetime symmetries, reflection positivity, and gauge invariance\label{subsec:spacemon}}

\subsubsection{Lorentz invariance}

Since the ZM operator $\bar{d}_{\pm}$ defined in Eq.~(\ref{eq:zmops})
creates a fermion in a $j\!=\!0$ state, one might intuitively
expect it to be Lorentz invariant. To see that this bears out, consider
a Lorentz transformation $\Lambda$ (rotation in Euclidean signature)
with $U(\Lambda)$ the corresponding $SU(2)_{\mathrm{rot}}$ action
on spinors. Since the monopole spinor harmonics $\mathcal{Y}_{\pm1,0,0}^{1/2}$
have total angular momentum $j\!=\!0$, they must satisfy the identity:\footnote{To explicitly verify this with the expressions for the harmonics in
Appendix~\ref{app:zmdrop} requires some care, for such expressions
are derived by solving the Euclidean Dirac equation in a fixed gauge.
The background gauge field $\mathcal{A}_{\mu}dx^{\mu}$ also transforms under
rotations, and one must make a subsequent gauge transformation to
bring back $\mathcal{Y}_{qjM}^{L}$ to its original form, as discussed
in Ref.~\cite{wu1976}.}
\begin{equation}
\mathcal{Y}_{\pm1,0,0}^{1/2}(\Lambda x)=U(\Lambda)\mathcal{Y}_{\pm1,0,0}^{1/2}(x),
\end{equation}
 using which we see that:
\begin{align}
\Lambda:\bar{d}_{\pm}(z) & \rightarrow\int d^{3}x\,\bar{\psi}(\Lambda^{-1}x)U^{\dagger}(\Lambda)\mathcal{Y}_{\pm 1,0,0}^{1/2}(x-z),\nonumber \\
 & =\int d^{3}x\,\bar{\psi}(x)U^{\dagger}(\Lambda)\mathcal{Y}_{\pm 1,0,0}^{1/2}\left(\Lambda(x-\Lambda^{-1}z)\right),\nonumber \\
 & =\int d^{3}x\,\bar{\psi}(x)\mathcal{Y}_{\pm 1,0,0}^{1/2}(x-\Lambda^{-1}z),\nonumber \\
 & =\bar{d}_{\pm}(\Lambda^{-1}z),\label{eq:lorscal}
\end{align}
as expected of a Lorentz scalar.

\subsubsection{$\mathcal{CRT}$}

Here we consider how the ZM operators transform under the discrete symmetries of continuum Euclidean QED$_3$: reflection $\mathcal{R}$, charge conjugation $\mathcal{C}$, and time reversal $\mathcal{T}$. These are to be distinguished from the microscopic symmetries of the projective symmetry group (PSG)~\cite{wen2007} for Dirac spin liquids on various lattices, to be discussed later in Sec.~\ref{sec:mqnum}.

We define reflections $\mathcal{R}_{\mu}$ to be in the $\mu$-coordinate. Let us consider
reflections $\mathcal{R}_{1}$ in the $x^{1}$ coordinate for concreteness. On spinors, this acts as $\psi\!\to\!\gamma_{1}\psi$ and $\bar{\psi}\!\to\!\bar{\psi}(-\gamma_1)$
so that a flavor-singlet mass $\bar{\psi}\psi$ breaks reflection symmetry. Under $\mathcal{R}_{1}$, the unit vector $\hat{\varphi}\!=\!-(\sin\varphi)\hat{x}\!+\!(\cos\varphi)\hat{y}\!\to\!-\hat{\varphi}$
so that the monopole background in the Wu-Yang gauge~\cite{wu1976} transforms as
$\mathcal{A}_{\mu}\!=\!(0,0,\mathcal{A}_{\varphi})\!\to\!(0,0,-\mathcal{A}_{\varphi})$,
which amounts to reversing the monopole charge $q\!\to\!-q$ as
expected of reflections. Explicitly, the monopole spinor harmonics obey:
\begin{equation}
\mathcal{Y}_{\pm 1,0,0}^{1/2}(\theta,\pi\!-\!\varphi)\!=\!(-\gamma_{1})\mathcal{Y}_{\mp 1,0,0}^{1/2}(\theta,\varphi),
\end{equation}
 under reflection $\mathcal{R}_{1}$ in the $x^{1}$-coordinate, so
that
\begin{align}\label{R1}
\mathcal{R}_{1}:\bar{d}_{\pm}(z) & \rightarrow\int d^{3}x\,\bar{\psi}(\mathcal{R}_{1}x)(-\gamma_{1})\mathcal{Y}_{\pm 1,0,0}^{1/2}(x-z),\nonumber \\
 & =\int d^{3}x\,\bar{\psi}(x)(-\gamma_{1})\mathcal{Y}_{\pm 1,0,0}^{1/2}\left(\mathcal{R}_{1}(x-\mathcal{R}_{1}z)\right),\nonumber \\
 & =\int d^{3}x\,\bar{\psi}(x)(-\gamma_{1})^{2}\mathcal{Y}_{\mp 1,0,0}^{1/2}(x-\mathcal{R}_{1}z),\nonumber \\
 & =\bar{d}_{\mp}(\mathcal{R}_{1}z).
\end{align}

Charge conjugation is a unitary symmetry that acts to send:
\begin{align}
\psi & \to-\gamma_{1}\psi^{*}=-\gamma_{1}\gamma_{3}\bar{\psi}^{\intercal}=(i\gamma_{2})\bar{\psi}^{\intercal},\nonumber \\
\bar{\psi} & \to\psi^{\intercal}(-\gamma_{1}\gamma_{3})=\psi^{\intercal}(i\gamma_{2}),
\end{align}
which flips the sign of the Dirac current, but not the mass. Using:
\begin{equation}
(i\gamma_{2})\mathcal{Y}_{\pm 1,0,0}^{1/2}(\theta,\varphi)=\pm\mathcal{Y}_{\mp 1,0,0}^{1/2}(\theta,\varphi)^{*},
\end{equation}
it is straightforward to verify that:
\begin{equation}\label{C-on-ZMs}
\mathcal{C}\colon \bar{d}_{\pm}(z)\to \pm d_{\mp}(z),\quad d_{\pm}(z)\to\mp \bar{d}_{\mp}(z).
\end{equation}

In Euclidean signature, time reversal as a spacetime symmetry behaves
identically to reflections~\cite{witten2016}, and is specifically
unitary. It can be defined as:
\begin{align}
\mathcal{T}  \colon&\psi(x)\to\gamma_{3}\psi(\mathcal{R}_{3}x),\nonumber \\
 & \bar{\psi}(x)\to\bar{\psi}(\mathcal{R}_{3}x)\gamma_{3},
\end{align}
where $\mathcal{R}_{3}$ is a reflection in the Euclidean time $(x^{3})$
coordinate. One can alternatively define a modified time-reversal operation $\mathcal{C}\mathcal{T}$
that also involves charge conjugation. On the ZM operators, 
\begin{align}
\mathcal{T}\colon \bar{d}_{\pm}(z) & \to \bar{d}_{\mp}(\mathcal{R}_{3}z),
\label{eq:bareTR}
\end{align}
using the fact that, under $\mathcal{R}_{3}$ reflections,
\begin{equation}
\mathcal{Y}_{\pm 1,0,0}^{1/2,(N)}(\pi-\theta,\varphi)=\gamma_{3}\mathcal{Y}_{\mp 1,0,0}^{1/2,(S)}(\theta,\varphi),\label{eq:r3Y}
\end{equation}
as one can verify from explicit expressions for the monopole harmonics.

\subsubsection{Reflection positivity\label{subsec:ref+}}

Reality of the real-time action (and thus unitarity of the corresponding quantum field theory) is
guaranteed by \emph{reflection positivity} $\vartheta(S)\!=\!S$ of the
Euclidean action~\cite{osterwalder1973,wetterich2011}. This is a form of complex
conjugation accompanied by a reversal of Euclidean time and an involution
of the Grassmann algebra. With our choices of coordinates and Dirac
matrices~\cite{shankar2021}, 
\begin{align}
\vartheta(\lambda\psi(x)) & \coloneqq\lambda^{*}\bar{\psi}(\mathcal{R}_{3}x)\gamma_{3},\nonumber \\
\vartheta(\lambda\bar{\psi}(x)) & \coloneqq\lambda^{*}\gamma_{3}\psi(\mathcal{R}_{3}x),\quad\lambda\!\in\!\mathbb{C},\nonumber \\
\vartheta(a_{\mu}(x)dx^{\mu}) & \coloneqq a_{\mu}(\mathcal{R}_{3}x)d(\mathcal{R}_{3}x)^{\mu},
\end{align}
and $\vartheta$ also reverses the order of Grassmann variables, e.g., $\vartheta(\psi_\alpha\psi_\beta\psi_\gamma)=\vartheta(\psi_\gamma)\vartheta(\psi_\beta)\vartheta(\psi_\alpha)$. For instance, one can check that the usual Berry phase
term $\int\bar{\psi}\gamma_{3}\partial_{\tau}\psi$ is reflection positive
using the definitions above. On the ZM operators $d_{\pm}$, we observe
that
\begin{align}
\vartheta(\bar{d}_{\pm}(z)) & =\int d^{3}x\,\mathcal{Y}_{\mp 1,0,0}^{1/2}(x-\mathcal{R}_{3}z)^{\dagger}\psi(x),\nonumber \\
 & =d_{\mp}(\mathcal{R}_{3}z),\label{eq:refpo}
\end{align}
where we have used the fact that reflections invert the monopole charge [see Eq.~\eqref{eq:r3Y}]. Together with the transformation $\vartheta(\sigma(z))=\sigma(\c{R}_3z)$ for the dual photon~\cite{shankar2021}, the transformation (\ref{eq:refpo}) ensures that the 't Hooft vertex (\ref{eq:finalvert})
is reflection positive, thereby implying the reality of the real-time
action or hermiticity of the Hamiltonian. It is important to note that reflection conjugation $\vartheta(d_q)$ replaces the notion of hermitian conjugation in Euclidean signature. In particular, we will \emph{define}:
\begin{equation}
	d_\pm^\dagger \coloneqq \vartheta(d_\pm) = \bar{d}_{\mp}.
\end{equation}

\subsubsection{Local gauge invariance}

We will prove invariance of the ZM operators under gauge transformations
with nonzero support on a sphere of fixed radius in $\mathbb{R}^{3}$.
By radial quantization, this suffices to prove gauge invariance in
general. The integrand of the expressions (\ref{eq:zmops}) should
be viewed as sections of a $U(1)$ bundle over punctured $\mathbb{R}^{3}$.
Charting a fixed sphere surrounding the monopole with ``northern''
(N) and ``southern'' (S) gauges \`a la Wu-Yang~\cite{wu1976}, it
is clear that $\psi$ should gauge transform identically to the spinor
harmonics $\mathcal{Y}_{q00}^{1/2}$:
\begin{align}
    \psi^{(N)}(x)&= e^{-iq\varphi} \psi^{(S)}(x),\nonumber \\
    \mathcal{Y}^{1/2,(N)}_{q,0,0}(x) &= e^{-iq\varphi} \mathcal{Y}^{1/2,(S)}_{q,0,0}(x),
\end{align}
for $\varphi$ the azimuthal
coordinate on $S^{2}\!\subset\!\mathbb{R}^{3}$. Since $\int d^3x\!=\!\int_{N\cap S}d^3x$, because the N and S poles are a set of measure zero in the integral, the mode operators transform as:
\begin{align}
\bar{d}_{q}&=  \int_{N\cap S} d^{3}x\,\bar{\psi}^{(N)}(x)\mathcal{Y}_{q,0,0}^{1/2,(N)}(x-z)\nonumber \\
&=  \int_{N\cap S} d^{3}x\,\bar{\psi}^{(S)}(x) (e^{-iq\varphi})^*e^{-iq\varphi}\mathcal{Y}_{q,0,0}^{1/2,(S)}(x-z)\nonumber \\
&= \int_{N\cap S} d^{3}x\,\bar{\psi}^{(S)}(x)\mathcal{Y}_{q,0,0}^{1/2,(S)}(x-z)\nonumber \\
&=  \bar{d}_{q}.
\end{align}
 A similar calculation shows invariance of $d_{q}$.

\subsection{Flavor symmetry}\label{subsec:flavor}

The global form of the symmetry group of compact QED$_{3}$ with $N_{f}$
flavors has been nicely summarized in Ref.~\cite{cordova2018}; let us review the necessary aspects here in our framework
and notation, for general $N_{f}$. The Lagrangian $\bar{\psi}\sld{D}\!\psi$
is invariant under $U(N_{f})$ rotations of the fermions, but the
center $U(1)$ is a gauge redundancy as it leaves spin operators invariant.
Moreover, it acts trivially on gauge-invariant fermion bilinears
such as $\bar{\psi}t^a\psi$. One might then conclude that the
symmetry group of the DSL is $PU(N_{f})\!\times\!U(1)_{\mathcal{M}}\cong\!PSU(N_{f})\!\times\!U(1)_{\mathcal{M}}$,
where $PSU(N_{f})\!\cong\!SU(N_f)/\mathbb{Z}_{N_f}$ and $U(1)_{\mathcal{M}}$ is the topological ``magnetic'' symmetry
corresponding to conservation of magnetic charge $\frac{1}{2\pi}\int f$
on any 2-cycle. However, monopole operators do not transform well as a representation of this group. The monopoles of minimal charge are precisely
the 't Hooft vertices calculated previously, and are of the form (for
general $N_{f}$):
\begin{equation}
e^{i\sigma}\Delta_{a_{1}\cdots a_{N_{f}/2}}d_{a_{1}}^{\dagger}\cdots d_{a_{N_{f}/2}}^{\dagger},
\end{equation}
with $\Delta$ totally antisymmetric in its $N_{f}/2$ indices. Under the
center of $SU(N_{f})$ generated by $e^{2\pi i/N_{f}}$, the vertex
transforms by an overall phase of $(e^{2\pi i/N_{f}})^{N_{f}/2}\!=\!-1$.
This is identical to a $\pi$ shift in $U(1)_{\mathcal{M}}$, which
implies the symmetry group is really:
\begin{equation}
\frac{SU(N_{f})\!\times\!U(1)_{\mathcal{M}}}{\mathbb{Z}_{N_{f}}},
\end{equation}
where the $\mathbb{Z}_{N_{f}}$ in the quotient is generated by:
\begin{equation}
(e^{2\pi i/N_{f}},-1)\in SU(N_{f})\!\times U(1)_{\mathcal{M}}.
\end{equation}
For $N_{f}\!=\!4$, the isomorphism $SU(4)/\mathbb{Z}_2\!\cong\!SO(6)$
can be used to equivalently write the symmetry group of the DSL as:
\begin{equation}
\frac{SO(6)\!\times\!U(1)_{\mathcal{M}}}{\mathbb{Z}_{2}},
\end{equation}
as concluded by Ref.~\cite{song2020}. 

A basis for the vector space of $q\!=\!\pm 1$ monopole operators can
then be constructed from the six antisymmetric generators of $\mathfrak{su}(4)$.
Doing so, we obtain three spin-singlet, valley-triplet monopoles:
\begin{align}\label{spin-singlet}
e^{iq\sigma}\bar{d}_{q}(-i\sigma_{2}\mu_{3})(\bar{d}_{q})^{\intercal} & \equiv \mathcal{V}_{1q},\nonumber \\
e^{iq\sigma}\bar{d}_{q}(\sigma_{2})(\bar{d}_{q})^{\intercal} & \equiv\mathcal{V}_{2q},\nonumber \\
e^{iq\sigma}\bar{d}_{q}(i\sigma_{2}\mu_{1})(\bar{d}_{q})^{\intercal} & \equiv\mathcal{V}_{3q},
\end{align}
and three spin-triplet, valley-singlet monopoles:
\begin{align}\label{spin-triplet}
e^{iq\sigma}\bar{d}_{q}(-\sigma_{3}\mu_{2})(\bar{d}_{q})^{\intercal} & \equiv\mathcal{S}_{1q},\nonumber \\
e^{iq\sigma}\bar{d}_{q}(i\mu_{2})(\bar{d}_{q})^{\intercal} & \equiv\mathcal{S}_{2q},\nonumber \\
e^{iq\sigma}\bar{d}_{q}(\sigma_{1}\mu_{2})(\bar{d}_{q})^{\intercal} & \equiv\mathcal{S}_{3q}.
\end{align}
It is straightforward to verify that these have the same spin/valley structure as the monopole operators defined in Refs.~\cite{song2019a,song2020}, up to some signs chosen
so that the six monopoles map to the standard basis of $\mathbb{C}^{6}$,
under the isomorphism from the $\bigwedge^2\mathbb{C}^4$ irrep
of $SU(4)$ to the vector irrep of $SO(6)$. In addition, there are operators reflection conjugate to those defined above:
\begin{equation}
	\c{V}_{iq}^\dagger \equiv \vartheta(\c{V}_{iq}),\qquad \c{S}_{iq}^\dagger \equiv \vartheta(\c{S}_{iq}),
\end{equation}
which we can use to construct the six operators
\begin{equation}
	\c{V}_{i} = \c{V}_{i+} + \c{V}_{i-}^\dagger,\quad
	\c{S}_{i} = \c{S}_{i+} + \c{S}_{i-}^\dagger,
\end{equation}
For example,
\begin{equation}
	\c{V}_2 = e^{i\sigma}(\bar{d}_{+}\sigma_2\bar{d}_{+}^{\,\intercal} + d_{+}^{\,\intercal}\sigma_2 d_{+}),
\end{equation}
is a monopole of definite magnetic charge (+1) that can create or annihilate pairs of spinons, as illustrated earlier in Fig.~\ref{fig:instzms}.

By examining the instanton-induced 't Hooft vertex (\ref{eq:finalvert}),
we observe that a choice of $\mathfrak{su}(4)$-adjoint mass proliferates
a linear combination of two of the six monopoles $\{\c{V}_i$, $\c{S}_i\}$. There are 15 such combinations, in correspondence with the 15 generators of $\mathfrak{su}(4)$.
As an example, the 't~Hooft vertex (\ref{eq:finalvert}) for a spin-Hall
mass $M_{30}\!=\!\bar{\psi}\sigma_{3}\psi$ can be written in the
above basis as
\begin{align}
\c{L}_{30} & = \c{S}_{1+}+i\c{S}_{2+}+\c{S}_{1-}-i\c{S}_{2-}+\mathrm{r.c.}\nonumber\\
& = \Re\c{S}_1 + \Im\c{S}_2,
\end{align}
defining $\Re\c{S}_i\!\equiv\!\c{S}_i\!+\!\c{S}_i^\dagger$ and $\Im\c{S}_i\!\equiv\!i(\c{S}_i\!-\!\c{S}_i^\dagger)$. Again, the adjoint $\S_i^{\dagger}$ of a monopole operator should
really be viewed in Euclidean signature as the ``reflection conjugate''
$\vartheta(\S_i)$ defined earlier in Sec.~\ref{subsec:ref+}. In this way we can find the
monopole operators spawned by all 15 adjoint masses, and we tabulate them in Table~\ref{tab:adjmon}.

\begin{table}
\begin{centering}
\begin{tabular}{|c|c|}
\hline 
Adjoint mass & Monopole proliferated\tabularnewline
\hline 
\hline 
$M_{01}$ & $\mathcal{V}_{3}+i\mathcal{V}_{2}+\mathrm{r.c.}$\tabularnewline
\hline 
$M_{02}$ & $\mathcal{V}_{3}+i\mathcal{V}_{1}+\mathrm{r.c.}$\tabularnewline
\hline 
$M_{03}$ & $-\mathcal{V}_{1}+i\mathcal{V}_{2}+\mathrm{r.c.}$\tabularnewline
\hline 
$M_{i1}$ & $\mathcal{S}_{i}-i\mathcal{V}_{1}+\mathrm{r.c.}$\tabularnewline
\hline 
$M_{i2}$ & $\mathcal{S}_{i}+i\mathcal{V}_{2}+\mathrm{r.c.}$\tabularnewline
\hline 
$M_{i3}$ & $\mathcal{S}_{i}-i\mathcal{V}_{3}+\mathrm{r.c.}$\tabularnewline
\hline 
$M_{i0}$ & $\mathcal{S}_{j}+i\mathcal{S}_{k}+\mathrm{r.c.}$ \tabularnewline
\hline 
\end{tabular}
\par\end{centering}
\caption{Monopoles proliferated by the 15 adjoint masses. ``r.c.'' denotes the reflection conjugate. In the last row, $(ijk)$ is an even permutation of $(123)$.}
\label{tab:adjmon}
\end{table}

\section{Monopole quantum numbers on bipartite lattices\label{sec:mqnum}}

It was observed in Refs.~\cite{song2019a,song2020} that there
exist orders on bipartite lattices whose microscopic symmetries are completely
captured by appropriate adjoint masses. Using such orders, we can
demand that the 't Hooft vertex induced by the given adjoint mass---i.e., the monopole proliferated by such a mass (Table~\ref{tab:adjmon})---must not break additional
symmetries, in order to fix its quantum numbers under certain lattice
symmetries. As we show below for the square lattice (Sec.~\ref{sec:square}) and the honeycomb lattice (Sec.~\ref{sec:honeycomb}), monopole quantum numbers on bipartite
lattices are reproduced accurately by this method. We expect that this is true for
any microscopic order that can be described in the continuum by condensing a
fermion bilinear. Conversely, there exist conventional orders whose broken symmetries are not fully captured by condensing a fermion bilinear. Examples include the $\bm{q}\!=\!0$
noncollinear magnetic states on the kagome lattice~\cite{hermele2008,lu2017}. Such orders have a $C_6$-breaking spin-ordering pattern which
is invisible to all 15 adjoint masses, but is captured by the spin-triplet
monopoles that serve as the correct order parameter for such states~\cite{song2019a,song2020}. (Precisely, it turns out that $C_{6}$
embeds into a $\mathbb{Z}_{3}^{\mathcal{M}}$ subgroup of $U(1)_{\mathcal{M}}$, as suspected initially in Ref.~\cite{hermele2008}.) On non-bipartite lattices, monopole proliferation breaks additional symmetries beyond those broken by the adjoint mass~\cite{song2019a}, thus our method for determining monopole quantum numbers does not apply to those cases.

\subsection{Square lattice}\label{sec:square}

On a square lattice, a DSL is obtained by coupling a staggered flux
mean-field state to $U(1)$ gauge fluctuations~\cite{hermele2005}.
We work with the gauge choice of Refs.~\cite{song2019a,song2020}
(but a different gamma matrix convention) which yields the following PSG action on the continuum Dirac spinor $\psi$:
\begin{align}
T_{x}\colon\psi &\to (-i\sigma_{2}\mu_{3})(i\gamma_{2})\bar{\psi}^{\intercal}, & \bar{\psi} &\to \psi^\intercal (i\gamma_2) (i\sigma_2\mu_3),\nonumber\\
T_{y}\colon\psi &\to (-i\sigma_{2}\mu_{1})(i\gamma_{2})\bar{\psi}^{\intercal}, & \bar{\psi} &\to \psi^\intercal (i\gamma_2) (i\sigma_2\mu_1),\nonumber\\
r_{x}\colon\psi &\to (\mu_{3}\gamma_{1}) \psi,
& \bar{\psi} &\to \bar{\psi} (-\gamma_1 \mu_3), \nonumber\\
C_{4s}\colon\psi &\to \frac{1}{\sqrt{2}}\sigma_{2}(i\mu_{2}\!-\!1)e^{-i\frac{\pi}{4}\gamma_{2}}(i\gamma_{2})\bar{\psi}^{\intercal}, & \bar{\psi} &\to \psi^\intercal e^{-i\frac{\pi}{4}\gamma_{2}}(i\gamma_{2}) \sigma_{2}(-i\mu_{2}\!-\!1)\frac{1}{\sqrt{2}},\nonumber \\
\Theta\colon\psi &\to Ki\mu_{2}(i\gamma_{2})\gamma_3\bar{\psi}^{\intercal}, & \bar{\psi} &\to \psi^\intercal (-i\mu_{2})(i\gamma_{2})\gamma_3 K,
\label{eq:psgsquare}
\end{align}
for $x$ and $y$ translations ($T_x$, $T_y$), reflections in the $x$ coordinate ($r_x$),
site-centered four-fold rotations ($C_{4s}$), and time reversal ($\Theta$), respectively, and $K$ denotes complex conjugation only on spin/valley matrices.

The embedding of the PSG into flavor (Sec.~\ref{subsec:flavor}) and spacetime (Sec.~\ref{subsec:spacemon}) symmetries in the
continuum completely fixes how the zero-mode part of the monopole
operators transform. However, the lattice symmetries also embed into
$U(1)_{\mathcal{M}}$, which acts on the bare monopole $\exp(i\sigma)$,
and this information is not present in the mean-field state from which
the above PSG is derived. The most general approach to calculating
this action, developed in Ref.~\cite{song2020}, is to consider
the Wannier limit, and the associated charge centers, of the spinon
insulator obtained on gapping the DSL with a given adjoint mass. In
this limit, the $U(1)_{\mathcal{M}}$ phase rotations of the monopole
under lattice symmetries are interpreted as Aharonov-Bohm phases.
For instance, a $C_{4s}$ action on a $q\!=\!+1$ monopole in an insulating
state with gauge charges $Q$ at lattice sites will yield a phase
$\exp(iQ\pi/2)$. 

\begin{table}
\begin{centering}
\begin{tabular}{|c|c|c|c|c|c|}
\hline 
 & $T_{x}$ & $T_{y}$ & $r_{x}$ & $C_{4s}$ & $\Theta$\tabularnewline
\hline 
\hline 
$M_{i0}$ & $-$ & $-$ & $-$ & $-$ & $-$\tabularnewline
\hline 
$M_{01}$ & $-$ & $+$ & $+$ & $M_{03}$ & $+$\tabularnewline
\hline 
$M_{03}$ & $+$ & $-$ & $-$ & $-M_{01}$ & $+$\tabularnewline
\hline 
$M_{02}$ & $+$ & $+$ & $+$ & $-$ & $-$\tabularnewline
\hline 
$M_{i1}$ & $+$ & $-$ & $+$ & $-M_{i3}$ & $+$\tabularnewline
\hline 
$M_{i3}$ & $-$ & $+$ & $-$ & $M_{i1}$ & $+$\tabularnewline
\hline 
$M_{i2}$ & $-$ & $-$ & $+$ & $+$ & $-$\tabularnewline
\hline 
\end{tabular}
\par\end{centering}
\caption{Transformation of the adjoint masses $M_{ij}\!=\!\bar{\psi}\sigma_{i}\mu_{j}\psi$
under the symmetries of the staggered-flux state on the square lattice.}
\label{tab:sqMij}
\end{table}

While no substitute for such rigorous microscopic arguments, we simply
note here that the existence of orders whose symmetries are fully
encapsulated by a fermion bilinear provides a simple means to compute
some, if not all, of the monopole quantum numbers. For example,
on the square lattice, the symmetries of N\'eel and valence-bond-solid (VBS) states are completely
encapsulated in the adjoint masses $M_{i2}$ and $M_{01/3}$, respectively
(see Table~\ref{tab:sqMij}). Let us demand that the monopoles proliferated by those masses (Table~\ref{tab:adjmon}) also remain invariant under the latter's symmetries.
As the VBS mass $M_{03}$ is $T_{x}$ invariant, we require that
the monopole $(-\Re\mathcal{V}_{1}\!+\!\Im\mathcal{V}_{2})$ also
be $T_{x}$ invariant. Likewise, $C_{4s}$ is a symmetry of the N\'eel mass
$M_{i2}$ which proliferates the monopole $\Re\mathcal{S}_{i}\!+\!\Im\mathcal{V}_{2}$. This means we can demand that $\Im\c{V}_2=i(\c{V}_2-\c{V}_2^\dag)$ be invariant under both $T_x$ and $C_{4s}$. However, from Eq.~(\ref{eq:psgsquare}) we see the corresponding PSG transformations involve charge conjugation $\psi\!\rightarrow\!(i\gamma_2)\bar{\psi}^{\intercal}$. Thus, the ZM operators $d_\pm$ will also undergo charge conjugation [Eq.~(\ref{C-on-ZMs})], and from Eq.~(\ref{spin-singlet}), $\c{V}_2$ will be mapped to its reflection conjugate $\c{V}_2^\dag$. The only way for $\Im\c{V}_2$ to remain invariant is thus to demand:
\begin{align}
T_{x}(\mathcal{V}_{2}) &=T_{x}(e^{i\sigma})(d_{-}^\intercal\sigma_{2}d_{-} + \bar{d}_{-}\sigma_{2}\bar{d}_{-}^{\,\intercal})\quad \stackrel{!}{=}-\mathcal{V}_{2}^{\dagger},\nonumber \\
C_{4s}(\mathcal{V}_{2}) &=C_{4s}(e^{i\sigma})(-d_{-}^\intercal\sigma_{2}d_{-}-\bar{d}_{-}\sigma_{2}\bar{d}_{-}^{\,\intercal}) \stackrel{!}{=}-\mathcal{V}_{2}^{\dagger},
\end{align}
which determines:
\begin{equation}
T_{x}(\sigma) =-\sigma+\pi,\qquad C_{4s}(\sigma) =-\sigma.\label{eq:txc4}
\end{equation}
The quantum numbers of $\sigma$ under other lattice symmetries can
be similarly calculated, but one can also exploit relational constraints
among the generators of the PSG (see Supplemental Material of Ref.~\cite{song2019a}).
Using that $T_{x}T_{y}$ and $\Theta T_{x}$ are symmetries of the
N\'eel order $M_{i2}$ leads to:
\begin{equation}
T_{y}(\sigma)=\Theta(\sigma)=-\sigma+\pi.\label{eq:tyTR}
\end{equation}

Finally, we look at reflections $r_{x}$ on the square lattice. Its embedding into the continuum symmetries involves the continuum reflection $\c{R}_1$, which has an action $\c{R}_1\colon \bar{d}_{\pm}\!\to\!\bar{d}_{\mp}$ on ZM operators [Eq.~\eqref{R1}]. On the monopole $\mathcal{V}_{2}$, we find that:
\begin{equation}
r_{x}(\mathcal{V}_{2}) =r_{x}(e^{i\sigma}) (\bar{d}_{-}\sigma_2\bar{d}_{-}^{\,\intercal} + d_{-}^{\,\intercal}\sigma_2 d_{-})
 = e^{i\theta_{r}}\mathcal{V}_{2}^\dagger.\label{eq:rx}
\end{equation}
As reflections are a symmetry of the N\'{e}el mass $M_{i2}$, we can demand invariance under $r_x$ of the $\Im \c{V}_2$ monopole it proliferates. This sets $\theta_r\!=\!\pi$ in Eq.~\eqref{eq:rx} and therefore
\begin{equation}
	r_{x} (\sigma) = -\sigma+\pi. \label{eq:rxsig}
\end{equation}

The set of equations (\ref{eq:txc4})-(\ref{eq:rxsig}) completely determines
the Berry phases of monopoles under the lattice symmetries (\ref{eq:psgsquare}).
The total action of these symmetries on monopole operators has been
summarized in Table~\ref{tab:sqmon}. We note that our results are identical
to the first four rows of Table 1 of Ref.~\cite{song2019a}. In particular, we also find that the monopole $\Im\mathcal{V}_{2}$ is trivial under all lattice symmetries. 
\begin{table}
\begin{centering}
\begin{tabular}{|c|c|c|c|c|c|}
\hline 
 & $T_{x}$ & $T_{y}$ & $r_{x}$ & $C_{4s}$ & $\Theta$\tabularnewline
\hline 
\hline 
$\mathcal{V}_{1}$ & $\mathcal{V}_{1}^{\dagger}$ & $-\mathcal{V}_{1}^{\dagger}$ & $-\mathcal{V}_{1}^\dagger$ & $-\mathcal{V}_{3}^{\dagger}$ & $\mathcal{V}_{1}^{\dagger}$\tabularnewline
\hline 
$\mathcal{V}_{2}$ & $-\mathcal{V}_{2}^{\dagger}$ & $-\mathcal{V}_{2}^{\dagger}$ & $-\mathcal{V}_{2}^\dagger$ & $-\mathcal{V}_{2}^{\dagger}$ & $-\mathcal{V}_{2}^{\dagger}$\tabularnewline
\hline 
$\mathcal{V}_{3}$ & $-\mathcal{V}_{3}^{\dagger}$ & $\mathcal{V}_{3}^{\dagger}$ & $\mathcal{V}_{3}^\dagger$ & $\mathcal{V}_{1}^{\dagger}$ & $\mathcal{V}_{3}^{\dagger}$\tabularnewline
\hline 
$\mathcal{S}_{i}$ & $-\mathcal{S}_{i}^{\dagger}$ & $-\mathcal{S}_{i}^{\dagger}$ & $\mathcal{S}_{i}^\dagger$ & $\mathcal{S}_{i}^{\dagger}$ & $-\mathcal{S}_{i}^{\dagger}$\tabularnewline
\hline 
\end{tabular}
\par\end{centering}
\caption{Monopole quantum numbers on the square lattice.}
\label{tab:sqmon}
\end{table}

We caution that one cannot expect the 't Hooft vertex to respect the
symmetries of the adjoint mass \emph{in general}, as demonstrated
by the results of Refs.~\cite{song2019a,song2020}. As an example,
consider the unconventional order:
\begin{equation}
M_{i3}\sim\sum_{\bm{r}}(-1)^{r_{x}}(\bm{S}_{\bm{r}}\times\bm{S}_{\bm{r}+\hat{y}})_i,\label{eq:M33}
\end{equation}
where the right-hand side is a spin operator on the square lattice with the same microscopic
symmetries as the fermion bilinear on the left-hand side~\cite{hermele2005,song2019a}. This equation suggests that $M_{i3}$
describes a spin-triplet VBS state invariant under $T_{y}$. However,
the monopole proliferated by $M_{i3}$ is $\mathcal{S}_{i}\!-\!i\mathcal{V}_{3}$.
By condensing this as $\ev{\mathcal{S}_{i}\!-\!i\mathcal{V}_{3}}\!=\!1\!-\!i$,
one observes that there is N\'eel order along $\sigma_{i}$ \emph{in
addition to} the order described by $M_{i3}$ (\ref{eq:M33}). This
follows from the fact that $\Re\mathcal{S}_{i}$ and $\Re\mathcal{V}_{3}$
have the symmetries of N\'eel order $M_{i2}$ and the triplet VBS order
(\ref{eq:M33}), respectively. The additional broken symmetries of
the N\'eel order are not visible to the adjoint mass $M_{i3}$ but are
captured by the associated 't Hooft vertex, which additionally breaks
$T_{y}$ and $\Theta$ symmetries. However, the above method
offers a quick way to compute quantum numbers when there exist orders
with symmetries completely encoded in a fermion bilinear, paradigmatic
examples being N\'eel and VBS orders.

\subsection{Honeycomb lattice}\label{sec:honeycomb}

On a honeycomb lattice, a parton mean-field Hamiltonian describing
uniform nearest-neighbor hopping has a relativistic dispersion with
gapless Dirac nodes at $\bm{K}_{\pm}\!=\!\pm\frac{4\pi}{3\sqrt{3}}\hat{y}$.
As is well-known, this model has a particle-hole symmetry which acts
trivially on the physical spin operators, and when combined with $U(1)$
gauge fluctuations yields an $SU(2)$ gauge theory (QCD$_{3}$) at
low energies~\cite{hermele2007}. However, the addition of longer-range hopping breaks
particle-hole symmetry and yields a DSL described by
CQED$_{3}$ in the infrared. Since the particle-hole symmetric state is
adiabatically connected to the DSL, we may calculate monopole quantum
numbers in the former for simplicity, and to make useful comparison
with the results of Refs.~\cite{song2019a,song2020}. Choosing a two-site (AB) unit cell on armchair graphene with Bravais lattice
vectors $\b{a}_{1/2}\!=\!(1/2,\pm\sqrt{3}/2)$, the PSG for the particle-hole
symmetric ansatz is:
\begin{align}
T_{1/2}\colon\psi &\to e^{-i2\pi\mu_{3}/3}\psi, & \bar{\psi} &\to \bar{\psi}e^{i2\pi\mu_{3}/3},\nonumber \\
C_{6}\colon\psi &\to -i\mu_{1}e^{-i2\pi\mu_{3}/3}e^{-i\pi\gamma_{1}/6}\psi, & \bar{\psi} &\to \bar{\psi}(ie^{i\pi\gamma_{1}/6}e^{i2\pi\mu_{3}/3}\mu_{1}),\nonumber \\
r_{x}\colon\psi &\to \mu_{2}\gamma_{3}\psi, & \bar{\psi} &\to \bar{\psi}(-\mu_{2}\gamma_{3})\nonumber \\
\Theta\colon\psi &\to K(i\sigma_{2}\mu_{2}\gamma_{3})\psi, & \bar{\psi} &\to \bar{\psi}(i\sigma_{2}\mu_{2}\gamma_{3})K,\label{eq:psghoney}
\end{align}
for (respectively) translations $T_{1/2}$ along $\b{a}_{1/2}$, plaquette-centered
six-fold rotations ($C_6$), reflections about the vertical axis through an
AB unit cell ($r_x$), and time reversal ($\Theta$). $K$ acts to complex conjugate only
within the spin-valley space (i.e., the matrices $\sigma_{i}$ and
$\mu_{i}$).

\begin{table}
\begin{centering}
\begin{tabular}{|c|c|c|c|c|}
\hline 
 & $T_{1/2}$ & $C_{6}$ & $r_{x}$ & $\Theta$\tabularnewline
\hline 
\hline 
$M_{i0}$ & $+$ & $+$ & $-$ & $+$\tabularnewline
\hline 
$M_{01}$ & $\alpha M_{01}\!+\!\beta M_{02}$ & $\alpha M_{01}\!+\!\beta M_{02}$ & $+$ & $+$\tabularnewline
\hline 
$M_{02}$ & $\alpha M_{02}\!-\!\beta M_{01}$ & $\!-\!\alpha M_{02}\!+\!\beta M_{01}$ & $-$ & $+$\tabularnewline
\hline 
$M_{03}$ & $+$ & $-$ & $+$ & $+$\tabularnewline
\hline 
$M_{i1}$ & $\alpha M_{i1}\!+\!\beta M_{i2}$ & $\alpha M_{i1}\!+\!\beta M_{i2}$ & $+$ & $-$\tabularnewline
\hline 
$M_{i2}$ & $\alpha M_{i2}\!-\!\beta M_{i1}$ & $\!-\!\alpha M_{i2}\!+\!\beta M_{i1}$ & $-$ & $-$\tabularnewline
\hline 
$M_{i3}$ & $+$ & $-$ & $+$ & $-$\tabularnewline
\hline 
\end{tabular}
\par\end{centering}
\caption{Transformation of the adjoint masses $M_{ij}\!=\!\bar{\psi}\sigma_{i}\mu_{j}\psi$
under the PSG (\ref{eq:psghoney}) on the honeycomb lattice, with $\alpha\!=\!\cos(\frac{2\pi}{3})$ and $\beta\!=\!\sin(\frac{2\pi}{3})$.}
\label{tab:honeyMij}
\end{table}

Similar to the square lattice in Sec.~\ref{sec:square}, we first tabulate
the transformation of the 15 adjoint masses $M_{ij}\!=\!\bar{\psi}\sigma_{i}\mu_{j}\psi$
under the above PSG. From Table~\ref{tab:honeyMij}, it is clear that
$M_{i3}$ encapsulates all the symmetries of N\'eel order on the honeycomb
lattice. We can then expect the associated proliferated monopole $\mathcal{S}_{i}\!-\!i\mathcal{V}_{3}$ (see Table~\ref{tab:adjmon}) to not break any additional symmetries, and thus demand:
\begin{equation}\label{V3honey}
T_{1/2}(\mathcal{V}_{3}) =T_{1/2}(e^{i\sigma})[\bar{d}_{+}(i\sigma_{2}\mu_{1})\bar{d}_{+}\!-\!d_{+}(i\sigma_{2}\mu_{1})d_{+}] \stackrel{!}{=}\mathcal{V}_{3},
\end{equation}
noting that to the difference of Eq.~(\ref{eq:psgsquare}), the PSG here does not involve charge conjugation. Equation~(\ref{V3honey}) implies that lattice translations act trivially on the dual photon. Turning to reflections, we similarly demand that $r_{x}(\mathcal{V}_{3})\!=\!e^{i\theta_{r}}\mathcal{V}_{3}^\dagger$ be equal to $\c{V}_3$, which leads to the action $r_x(\sigma)\!=\!-\sigma$ with no Berry phase.

\begin{figure}[h]
\begin{centering}
\includegraphics[width=0.8\textwidth,height=1\textwidth,keepaspectratio]{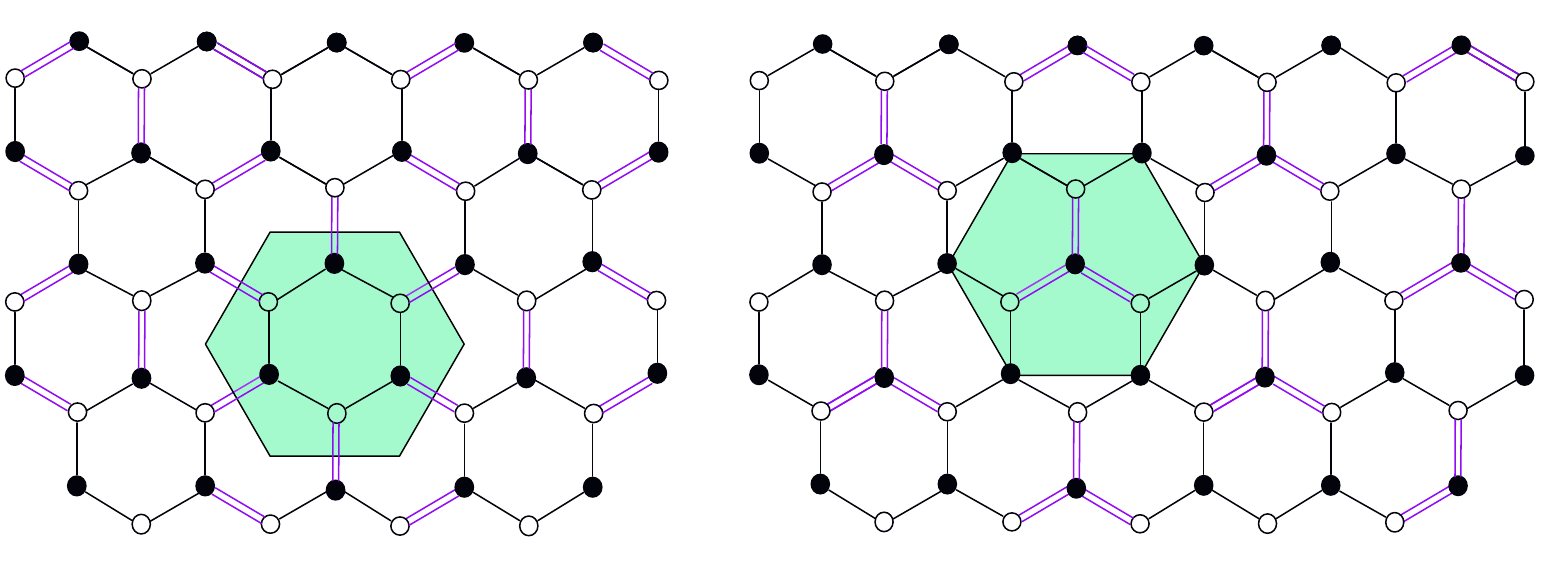}
\par\end{centering}
\caption{Kekul\'e-O (left) and Kekul\'e-Y (right) patterns on the honeycomb lattice.}
\label{fig:kekyo}
\end{figure}

Similarly, $M_{01}$ and $M_{02}$ account for all symmetries of the
Kekul\'e-O and Kekul\'e-Y VBS states, respectively (Fig.~\ref{fig:kekyo}).
We can use the time-reversal invariance of these orders to demand invariance of the monopole $\Re \c{V}_3$ that both proliferate:
\begin{equation}
\Theta(\mathcal{V}_{3})=\Theta(e^{i\sigma})[\bar{d}_{-}(i\sigma_{2}\mu_{1})\bar{d}_{-}^{\,\intercal}\!-\!{d}_{-}^\intercal(i\sigma_{2}\mu_{1}){d}_{-}] =e^{i\theta_{\Theta}}\mathcal{V}_{3}^\dagger \stackrel{!}{=}\mathcal{V}_{3}^\dagger,
\end{equation}
and so $\theta_{\Theta}\!=\!0$. Finally, to compute
quantum numbers under $C_{6}$, we may use the $C_{6}\Theta$
symmetry of the N\'eel mass $M_{i3}$ to demand invariance of $\Im\c{V}_3$:
\begin{align}
(C_{6}\!\circ\!\Theta)(i(\c{V}_{3}-\c{V}_3^\dagger))&=-iC_{6}(\c{V}_3-\c{V}_3^\dagger),\nonumber\\
&= -ie^{-i\theta_6}(\c{V}_3-\c{V}_3^\dagger),\nonumber\\
&\stackrel{!}{=} i(\c{V}_{3}-\c{V}_3^\dagger),
\end{align}
which requires $\theta_6\!=\!\pi$. Collecting our results,
the lattice symmetries act on the dual photon as follows: 
\begin{align}
T_{1/2}(\sigma) & =\sigma,\nonumber \\
r_{x}(\sigma) & =-\sigma,\nonumber \\
\Theta(\sigma) & =-\sigma,\nonumber \\
C_{6}(\sigma) & =\sigma+\pi,
\end{align}
from which transformations of all six monopole operators can be determined. These results are summarized in Table~\ref{tab:honeymon}, and agree with the results in Table 1 of Ref.~\cite{song2019a} for the honeycomb lattice. The monopole $\Re\mathcal{V}_{3}$
is trivial under all lattice symmetries and is thus a symmetry-allowed
perturbation to the DSL.

\begin{table}[h]
\begin{centering}
\begin{tabular}{|c|c|c|c|c|}
\hline 
 & $T_{1/2}$ & $r_{x}$ & $C_{6}$ & $\Theta$\tabularnewline
\hline 
\hline 
$\mathcal{V}_{1}$ & $\alpha\mathcal{V}_{1}\!-\!\beta\mathcal{V}_{2}$ & $\mathcal{V}_{1}^{\dagger}$ & $\!-\!\alpha\mathcal{V}_{1}\!+\!\beta\mathcal{V}_{2}$ & $\mathcal{V}_{1}^{\dagger}$\tabularnewline
\hline 
$\mathcal{V}_{2}$ & $\alpha\mathcal{V}_{2}\!+\!\beta\mathcal{V}_{1}$ & $-\mathcal{V}_{2}^\dagger$ & $\alpha\mathcal{V}_{2}\!+\!\beta\mathcal{V}_{1}$ & $\mathcal{V}_{2}^\dagger$\tabularnewline
\hline 
$\mathcal{V}_{3}$ & $\mathcal{V}_{3}$ & $\mathcal{V}_{3}^\dagger$ & $\mathcal{V}_{3}$ & $\mathcal{V}_{3}^\dagger$\tabularnewline
\hline 
$\mathcal{S}_{i}$ & $\mathcal{S}_{i}$ & $-\mathcal{S}_{i}^\dagger$ & $-\mathcal{S}_{i}^{\dagger}$ & $-\mathcal{S}_{i}^\dagger$\tabularnewline
\hline 
\end{tabular}
\par\end{centering}
\caption{Monopole quantum numbers on the honeycomb lattice, with $\alpha\!=\!\cos(\frac{2\pi}{3})$ and $\beta\!=\!\sin(\frac{2\pi}{3})$.}
\label{tab:honeymon}
\end{table}

\section{Conclusion\label{sec:concl}}

In summary, we have constructed monopole operators in the DSL directly on $\mathbb{R}^{3}$ without assuming conformal invariance,
and have computed their quantum numbers under lattice symmetries on
the square and honeycomb (bipartite) lattices. The first task was
accomplished by first deforming the DSL with a choice of an $\mathfrak{su}(4)$-valued
fermion mass. This was shown to lead to ZMs of the Euclidean
Dirac operator exponentially bound to monopole-instantons. The interpretation
of these ZMs in the Hamiltonian framework and their relation
to zero-energy modes was also discussed. We then showed that resumming
a semiclassical instanton gas in the presence of such ZMs leads to an instanton-induced
effective interaction, designated as the 't Hooft vertex in analogy with
a similar effect in QCD$_{4}$. By introducing ZM creation/annihilation
operators, we then identified this vertex as a linear combination
of two of six possible monopole operators in the DSL, previously constructed
in radially-quantized conformal CQED$_{3}$.

Our next result involved an analysis of the effects of lattice symmetries
in specific microscopic realizations of the DSL. By recognizing the
existence of orders on bipartite lattices with symmetries fully encapsulated
in a specific fermion bilinear, we were able to compute quantum numbers
of all monopoles under symmetries of the DSL on square and honeycomb
lattices. Specifically, from a symmetry standpoint, N\'eel and VBS orders on these lattices could
be described in the continuum by \emph{either} appropriate fermion
bilinears \emph{or} monopole operators (although monopole proliferation is necessary to confine spinons). By knowing the 't Hooft vertex
associated to a given bilinear, we could then demand that the former
\emph{not} break additional symmetries of the DSL to fix the lattice symmetry
action on monopoles. N\'eel and VBS orders on the square and honeycomb
lattices together possess enough unbroken lattice symmetries to fully
determine the transformations of all monopole operators. In particular,
our results for the ``Berry phase'' of monopoles, arising from the
embedding of the lattice symmetries into the magnetic symmetry $U(1)_{\mathcal{M}}$
of the dual photon, were shown to be consistent with the more general
Wannier center calculations of Refs.~\cite{song2019a,song2020}.
On both square and honeycomb lattices, we showed the existence of
a monopole transforming trivially under all lattice symmetries, and thus an allowed
perturbation to the DSL on these lattices likely to lead to its instability. 

\section*{Acknowledgements}
We thank S. Dey for useful discussions. G.S. was supported by the Golden Bell Jar Graduate Scholarship in Physics and an Alberta Graduate Excellence Scholarship. J.M. was supported by NSERC Discovery Grants RGPIN-2020-06999 and RGPAS-2020-00064; the Canada Research Chair (CRC) Program; the Government of Alberta's Major Innovation Fund (MIF); and the Pacific Institute for the Mathematical Sciences (PIMS) Collaborative Research Group program.

\begin{appendix}

\section{Zero modes of Euclidean Dirac operators\label{app:zmdrop}}

Since an index theorem for Dirac operators with abelian gauge fields
on odd-dimensional noncompact manifolds has not been established,
we resort to an explicit calculation of zero modes.

A charge-$q\!\in\!\mathbb{Z}$ monopole-instanton can be described by a Wu-Yang
connection, 
\begin{equation}
\mathcal{A}_{q}=\begin{cases}
-\frac{q}{2r\sin\theta}(\cos\theta-1)\hat{\b{\varphi}}, & \theta\!\in\!(0,\pi/2),\\
-\frac{q}{2r\sin\theta}(\cos\theta+1)\hat{\b{\varphi}}, & \theta\!\in\!(\pi/2,\pi),
\end{cases}
\end{equation}
in spherical coordinates with an orthonormal frame $(\hat{\b{r}},\hat{\b{\theta}},\hat{\b{\phi}}).$
We will explicitly solve for the zero modes of the Euclidean (non-self-adjoint) Dirac
operator in an instanton background,
\begin{equation}
\mathfrak{D}_{aq}=\slashed{\partial}-i\slashed{\mathcal{A}}_{q}+mt^{a},
\end{equation}
where $t^{a}\!\in\!\mathfrak{su}(4)$. Using the fact that $(\bm{\gamma}\!\cdot\!\hat{\bm{r}})^{2}\!\equiv\!\gamma_{r}^{2}\!=\!1$,
with $\bm{\gamma}\!=\!(\gamma_{1},\gamma_{2},\gamma_{3})$ the Pauli
vector, the Dirac operator can be rewritten as~\cite{shankar2021,borokhov2002}:
\begin{equation}
\gamma_{r}^{2}\,\mathfrak{D}_{aq}=\gamma_{r}\left(\partial_{r}-\frac{1}{r}\bm{\gamma}\!\cdot\!\bm{L}-\frac{q}{2r}\gamma_{r}\right)+mt^{a},
\end{equation}
 where:
\begin{equation}
\bm{L}=\bm{r}\!\times\!(\bm{p}\!-\!\bm{a})-\frac{q}{2}\hat{\bm{r}},
\end{equation}
is the conserved angular momentum in a monopole field. Defining the
total angular momentum:
\begin{equation}
\bm{J}=\bm{L}+\frac{1}{2}\bm{\gamma},
\end{equation}
 the Dirac operator takes the form:
\begin{equation}
\mathfrak{D}_{aq}=\gamma_{r}\left[\partial_{r}-\frac{1}{r}(\b{J}^{2}\!-\!\b{L}^{2}\!-\!\frac{3}{4})-\frac{q}{2r}\gamma_{r}\right]+mt^{a}.
\end{equation}
 To find the eigenfunctions, note that $\b{J}^{2},\,J_{z},\,t^{a}$ and
$\mathfrak{D}_{aq}$ commute. This prompts an eigenfunction ansatz:
\begin{equation}
u_{jM}^{qi}\!=\!R(r)\mathcal{Y}_{qjM}^{j+1/2}(\theta,\varphi)\ket{i}_a\!+\!S(r)\mathcal{Y}_{qjM}^{j-1/2}(\theta,\varphi)\ket{i}_{a},\label{eq:ans}
\end{equation}
where $\ket{i}_{a}$ is one of the four eigenvectors of $t^{a}$ with
eigenvalue $(-1)^{i}$, and $\mathcal{Y}_{qjM}^{L}$ are monopole
spinor harmonics defined in Appendix A of Ref.~\cite{shankar2021},
and also in \cite{borokhov2002}. Their necessary properties are
summarized as follows:
\begin{align}
\b{J}^{2}\mathcal{Y}_{qjM}^{L} & =j(j+1)\mathcal{Y}_{qjM}^{L},\nonumber \\
\b{L}^{2}\mathcal{Y}_{qjM}^{L} & =L(L+1)\mathcal{Y}_{qjM}^{L},\nonumber \\
J_{z}\mathcal{Y}_{qjM}^{L} & =M\mathcal{Y}_{qjM}^{L},\nonumber \\
\gamma_{r}\mathcal{Y}_{qjM}^{j\pm1/2} & =a_{\pm}\mathcal{Y}_{qjM}^{j+1/2}+b_{\pm}\mathcal{Y}_{qjM}^{j-1/2},
\end{align}
 where: 
\begin{align}
j & \in\left\{\frac{\abs{q}}{2}\!-\!\frac{1}{2},\frac{\abs{q}}{2}\!+\!\frac{1}{2},\ldots\right\},\,\,(j\!>\!0),\nonumber \\
M & \in\{-j,-j+1,\ldots,j\},\nonumber \\
L & \in\left\{j\!-\!\frac{1}{2},j\!+\!\frac{1}{2}\right\},\,\,\left(L\!\geq\!\frac{\abs{q}}{2}\right),\nonumber \\
a_{+} & =-b_{-}=\frac{q}{2j+1},\quad a_{-}=b_{+}=-\frac{\sqrt{(2j+1)^{2}-q^{2}}}{2j+1}.
\end{align}
The condition $j\!>\!0$ implies $j\!=\!(\abs{q}\!-\!1)/2$ is excluded
when $q\!=\!0$, and the condition $L\!\geq\!\abs{q}$ requires that
$L\!=\!j\!-\!1/2$ be excluded when $j\!=\!(\abs{q}\!-\!1)/2$. Therefore,
for a fixed $q$, the lowest angular momentum states with $j\!=\!(\abs{q}\!-\!1)/2$
have $S(r)\!=\!0$ in the ansatz (\ref{eq:ans}). As we shall now
show, these states are zero modes. The zero mode equation for $\mathfrak{D}_{aq}$
then separates to:
\begin{equation}
\left(\partial_{r}R+\frac{1}{r}R+\sgn(q)m_{i}\right)R(r)=0,
\end{equation}
where $m_{i}\!=\!(-1)^{i}m$ corresponding to $\ket{i}_{a}$, the
$SU(4)$ part of the zero mode. Solving for the radial function $R(r)$,
the zero modes can be written as:
\begin{align}
u_{\nicefrac{(q-1)}{2},M}^{qi} & =R^{qi}(r)\mathcal{Y}_{q,\nicefrac{(q-1)}{2},M}^{q}(\theta,\varphi)\ket{i}_{a},\nonumber \\
 & =\frac{\sqrt{2m}}{r}e^{-\sgn(q)(-1)^{i}mr}\mathcal{Y}_{q,\nicefrac{(q-1)}{2},M}^{q}\ket{i}_{a}.
 \label{eq:appzm}
\end{align}
For a fixed monopole charge $q$ and $\mathfrak{su}(4)$ mass $mt^{a}$,
it is clear that there are $2q\!\times\!\frac{N_{f}}{2}\!=\!qN_{f}$
linearly independent \emph{normalizable} zero modes.\footnote{It is assumed that $N_{f}$ is even, so that there is no parity anomaly.
In the case of $N_{f}$ odd, the non-anomalous theory has a half-integral
Chern-Simons term that can be regarded as the result of integrating
out an extra Dirac fermion.} We have utilized the fact that for $j\!=\!(q\!-\!1)/2$, the quantum
number $M$ ranges over the $2q$ values $-j,\ldots,j$, and that a given
sign of $q$ results in precisely two of the four eigenvectors $\ket{i=1,2,3,4}_{a}$
contributing normalizable ZMs.

It is also important to consider zero modes of the adjoint Dirac operator $\mathfrak{D}^{\dagger}$,
for the Dirac action can be rewritten after an integration by parts
and throwing away boundary terms as:
\begin{align}
S_{f} & =\int d^{3}x\,\bar{\psi}(\sld{\partial}-i\sld{a}+mt^{a})\psi,\nonumber \\
 & =\int d^{3}x\,[(-\sld{\partial}+i\sld{a}+mt^{a})\bar{\psi}^{\dagger}]^{\dagger}\psi,
\end{align}
where it is to be remembered that $\bar{\psi}$ and $\psi$ are independent
variables in the Euclidean path integral, unrelated by any notion
of complex conjugation. Repeating the calculation above leads to the
zero modes:
\begin{align}
v_{\nicefrac{(q-1)}{2},M}^{qi} & =R^{qi}(r)\mathcal{Y}_{q,\nicefrac{(q-1)}{2},M}^{q}(\theta,\varphi)\ket{i}_{a},\nonumber \\
 & =\frac{\sqrt{2m}}{r}e^{\sgn(q)(-1)^{i}mr}\mathcal{Y}_{q,\nicefrac{(q-1)}{2},M}^{q}\ket{i}_{a},
\end{align}
 where again, for a given $q$, $i$ must be chosen to ensure normalizability. 
 
For reference, we give expressions for the monopole spinor harmonics, also given in Appendix A of Ref.~\cite{shankar2021}:
 
 \begin{align}
\mathcal{Y}_{q,j,m}^{j\!-\!1/2}(\theta,\varphi)= & \frac{1}{\sqrt{2j}}\begin{pmatrix}\sqrt{j\!+\!m_{j}}Y_{q,j\!-\!\frac{1}{2},m\!-\!\frac{1}{2}}\\
\sqrt{j\!-\!m}Y_{q,j\!-\!\frac{1}{2},m\!+\!\frac{1}{2}}
\end{pmatrix},\nonumber \\
\mathcal{Y}_{q,j,m}^{j\!+\!1/2}(\theta,\varphi)= & \frac{1}{\sqrt{2j\!+\!2}}\begin{pmatrix}-\sqrt{j\!-\!m_{j}\!+\!1}Y_{q,j\!+\!\frac{1}{2},m\!-\!\frac{1}{2}}\\
\sqrt{j\!+\!m\!+\!1}Y_{q,j\!+\!\frac{1}{2},m\!+\!\frac{1}{2}}
\end{pmatrix}.\label{eq:spinMH}
\end{align}
The monopole harmonics $Y_{qLM}$ are defined in terms of the Wigner $D$-matrices $D^J_{MM'}(\alpha,\beta,\gamma)$ \cite{stone1989,wu1976}. In the northern chart on a sphere that surrounds the monopole-instanton,
\begin{equation}
Y_{q,L,M}(\theta_{N},\varphi)=\sqrt{\frac{2L+1}{4\pi}}\left[D_{M,-\nicefrac{q}{2}}^{L}(\varphi,\theta,-\varphi)\right]^{*},
\end{equation}
where $\theta_{N}\!\in\![0,\pi).$ The southern versions (which are valid on
the south pole) can be obtained via a gauge transformation on the overlapping region between northern and southern charts:
\begin{equation}
Y_{q,L,M}(\theta_{S},\varphi)=e^{-i2q\varphi}Y_{q,L,M}(\theta_{N},\varphi).
\end{equation}

From the above formula, the first two $q\!=\!1$ harmonics are
given by 
\begin{align}
Y_{1,\frac{1}{2},\frac{1}{2}}(\theta_{N},\varphi) & =-\frac{1}{\sqrt{2\pi}}e^{i\varphi}\sin\frac{\theta}{2},\nonumber \\
Y_{1,\frac{1}{2},-\frac{1}{2}}(\theta_{N},\varphi) & =\frac{1}{\sqrt{2\pi}}\cos\frac{\theta}{2},
\end{align}
in the northern chart. Their analogs on the southern chart are obtained from the gauge transformation
$\exp(-i\varphi)$. 

For $q\!=\!-1$, the first two harmonics on the northern chart are 
\begin{align}
Y_{-1,\frac{1}{2},\frac{1}{2}}(\theta_{N},\varphi) & =\frac{1}{\sqrt{2\pi}}\cos\frac{\theta}{2},\nonumber \\
Y_{-1,\frac{1}{2},-\frac{1}{2}}(\theta_{N},\varphi) & =\frac{1}{\sqrt{2\pi}}e^{-i\varphi}\sin\frac{\theta}{2},
\end{align}
with their versions in the southern chart now obtained from the gauge transformation
$\exp(i\varphi)$. 

\bigskip

\section{Real-space Dirac propagator\label{sec:xdirac}}

With the Lagrangian $\bar{\psi}(\slashed{\partial}\!+\!m)\psi$ where
$m$ is a signed quantity, the free Dirac propagator on $\mathbb{R}^{3}$ is:
\begin{align}
G_{f}(x) & =\int\frac{d^{3}k}{(2\pi)^{3}}e^{ikx}\frac{i\slashed{k}-m}{k^{2}+m^{2}},\nonumber \\
 & =(\slashed{\partial}-m)\int\frac{d^{3}k}{(2\pi)^{3}}\frac{e^{ikx}}{k^{2}+m^{2}},\nonumber \\
 & =(\gamma_{r}\partial_{r}-m)\frac{e^{-\abs{m}r}}{4\pi r},\nonumber \\
 & =-\frac{e^{-\abs{m}r}(1+\abs{m}r)}{4\pi r^{2}}\gamma_{r}-\frac{me^{-\abs{m}r}}{4\pi r}.
\end{align}

\end{appendix}



\bibliography{dsl.bib}

\nolinenumbers

\end{document}